\documentclass[physrev,graphicx,twocolumn,superscriptaddress,reprint]{revtex4-2}
\usepackage{vacuum_arc_magnetic_field}

\usepackage[pdftex]{hyperref}

\usepackage{academicons}
\usepackage{orcidlink}

\draft 

\begin{document}


\title{Influence of an external static magnetic field on prebreakdown electron emission and heating} 



\newcommand{\helsinki}{
Helsinki Institute of Physics and Department of Physics, P.O. Box 43 (Pietari Kalmin Katu 2), FI-00014 University of Helsinki, Finland
}
\newcommand{\tartu}{
Institute of Technology, University of Tartu, Nooruse 1, 50411 Tartu, Estonia
}
\newcommand{\cern}{
European Organization for Nuclear Research (CERN), CH-1211 Geneva 23, Switzerland
}
\author{Roni Koitermaa \orcidlink{0000-0001-9814-7358}}
\email[]{roni.koitermaa@helsinki.fi}
\email[]{roni.koitermaa@iki.fi}
\affiliation{\helsinki}
\affiliation{\tartu}

\author{Marzhan Toktaganova \orcidlink{0009-0001-9278-4214}}
\affiliation{\helsinki}

\author{Andreas Kyritsakis \orcidlink{0000-0002-4334-5450}}
\affiliation{\tartu}

\author{Tauno Tiirats \orcidlink{0000-0002-1256-7392}}
\affiliation{\tartu}

\author{Alexej Grudiev}
\affiliation{\cern}

\author{Veronika Zadin \orcidlink{0000-0003-0590-2583}}
\email[]{veronika.zadin@ut.ee}
\affiliation{\tartu}

\author{Flyura Djurabekova \orcidlink{0000-0002-5828-200X}}
\email[]{flyura.djurabekova@helsinki.fi}
\affiliation{\helsinki}


\date{\today}

\begin{abstract}
  High magnetic fields can increase the occurrence of vacuum arcing, suggesting that both electric and magnetic fields can play a role in the vacuum arcing process. The mechanism of vacuum arcing in high magnetic fields is believed to involve both the cathode and the anode, with the cathode serving as the originator of field-emitting nanoprotrusions or tips, while the anode serves a secondary role. Significant heating of the anode surface can be achieved by magnetic focusing of the emitted electron beam, leading to increased heat flux due to greater current density. We simulated the emitted electron beam in different configurations of the electric and magnetic fields using the particle-in-cell (PIC) and finite element methods (FEM). The heating caused by the impacting electron beam was simulated for magnetic fields ranging from 0~T to 30~T. The directions of the electric and magnetic fields were found to play a major role in the focusing of the electron beam. We found that a sufficient temperature increase on the anode surface for evaporation can be reached at magnetic fields on the order of 10--30~T, suggesting the possibility of plasma initiation on the anode side.
\end{abstract}


\maketitle 



%
%

%

\section{Introduction}\label{sec:introduction}

Vacuum arcing presents technical challenges in applications where high electric fields are used, such as electron sources, particle accelerators and vacuum interrupters. The origin and mechanics of these breakdowns has been the subject of numerous studies, but a complete picture of the process is yet to be obtained.

While both electric and magnetic fields are present during the vacuum arcing process, it has been determined that the electric field dominates over the magnetic field as the main arc initiation factor. The cathode is typically considered the main originator of DC vacuum arcs, while the anode plays a lesser role~\cite{zho19,and08,zho20, zho21}. This cathodic arc initiation has been proposed to be due to the self-heating and thermal runaway of microscopic protrusions on the cathode as a result of field-induced electron emission, which provides the influx of neutrals for plasma formation.

The roles of the cathode and anode could be reversed under certain conditions~\cite{uts67, cha67}, especially in the presence of a strong static or quasi-static magnetic field. In that case, a different arc initiation mechanism might provide the initial source of neutrals for the plasma onset. Such conditions may be present in particle accelerators, such as the proposed muon collider~\cite{muc24}. While accelerators use RF cavities where there is no fixed cathode and anode, the DC plate-to-plate electrode case serves as a useful simplified model where vacuum breakdown mechanisms can be studied and distinguished~\cite{des09}. Design studies foresee peak fields of 17~T~\cite{zhu25} in the rectilinear cooling channel and up to 30~T~\cite{bot24} in the final-cooling stage. At such high magnetic fields, vacuum arcing can become a major design consideration.

Design studies envisage a muon collider operating at roughly $10 \us{TeV}$ center-of-mass energy. However, delivering and sustaining such intense beams presents significant technical challenges. Ionization cooling is indispensable for a multi-TeV muon collider because muon beams emerge from the target with high six-dimensional emittance and must be shrunk by $\geq 10^5$ before most muons decay ($\tau_{\mu} = 2.2 \us{\textmu s}$)~\cite{muc24}. Each cooling cell combines three colinear elements~\cite{zhu25}: low-Z absorber, normal conducting RF cavities and superconducting solenoids. The RF cavities are typically pillbox-type single-cell structures with large apertures (e.g. $\sim 28 \us{cm}$ diameter iris at $352 \us{MHz}$) run at high vacuum levels ($\sim 10^{-7}\us{Torr}$)~\cite{zhu25,chu14}. These cavities sit inside the solenoid bore, so the axial magnetic field is parallel to the accelerating electric field. That overlap is vital for cooling but drives vacuum breakdown. Field-emitted electrons are forced by the solenoid into narrow beamlets that strike and locally melt the opposite wall. Measurements at the MuCool Test Area showed that an $805\us{MHz}$ copper cavity that held $40 \us{MV}/\un{m}$ at $0\us{T}$ was limited to $\sim 12 \us{MV}/\un{m}$ at $3\us{T}$~\cite{bow20,pal09}. MICE’s $201\us{MHz}$ cavities achieved $16 \us{MV}/\un{m}$ at $3\us{T}$ only after extensive conditioning~\cite{bow12}. Because later cooling stages demand both higher gradients ($\ge 25 \us{MV}/\un{m}$) and higher magnetic fields ($>10\us{T}$), the probability of vacuum breakdown grows precisely where maximum performance is required.

Current theories on the origin of vacuum arcs give the appearance of nanoprotrusions or tips on the metal surface as the origin of local field enhancement and field emission~\cite{wue25, kyr18, ves20}. Although the mechanism of tip growth remains unclear despite extensive study~\cite{poh13, zad14, jan20, kim22}, we assume the presence of these tips on the cathode surface. Strength of field emission is dependent on factors such as geometric field enhancement. This electron emission can initiate an arc via two distinct mechanisms, which we shall call ``anodic'' and ``cathodic'' following Utsumi's nomenclature~\cite{uts67}. Although in RF structures, there are no anode and cathode, both distinct mechanisms might appear depending on the conditions. In the cathodic mechanism, which has been extensively studied elsewhere~\cite{kyr18, ves20, koi24}, field emission causes self-heating of the protrusion, leading to evaporation of neutral atoms from the cathode surface, which ionize to form the arc plasma. In the anodic mechanism, the electrons are accelerated by the field and cause intense heating at the electrode that captures the electrons (the anode in DC), initiating the arc there. In this study, we aim to investigate the possibility of evaporation of the anode material as a result of focused emitted electron beams. If a sufficiently dense neutral vapor can accumulate above the anode surface, this may provide the preconditions for anode-initiated breakdown. If these conditions are met and ions can form on the anode side, they would be accelerated from the anode to the cathode, potentially causing heating, evaporation and sputtering.

High axial magnetic fields have been observed to increase arc occurrence~\cite{mor05}. In the presence of an external magnetic field, the emitted electron beam can be significantly focused, as supported by simulations and experiments~\cite{pal09, bow20, urb24}. This focusing depends on the characteristics of electron emission (tip geometry and other factors), as well as the magnitude and direction of the electric and magnetic fields. This focusing of the electron beam could lead to significant heating and potential arc initiation on the anode side. We use multi-physics simulations of both the electron beam focusing characteristics and the anode heating to study this process, using the DC case as a simplified model to investigate the underlying physical mechanisms.

\section{Methods}\label{sec:methods}

\fig{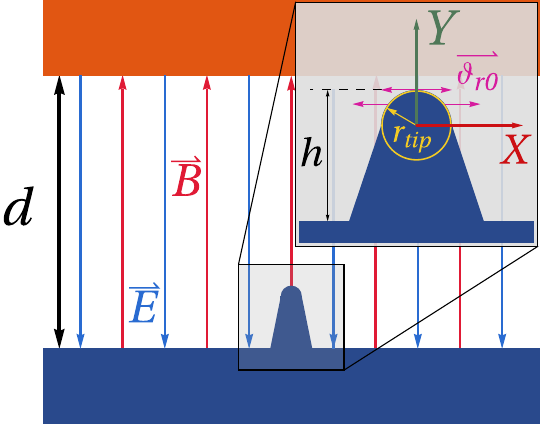}{The geometry of the system. The parallel electrodes (the cathode is shown in blue and the anode is in red) are separated by the gap length $d$. The blue and red arrows show the directions of electric $\vb{E}$ and magnetic $\vb{B}$ fields. A nanotip is placed on the surface of the cathode. The inset shows the height $h$ and the curvature radius $r_\x{tip}$ of the nanotip, the $X$ and $Y$ indicate the Cartesian coordinates used in the calculations. The initial velocity of the electrons $v_{r0}$ is due to the field enhancement around the nanotip.\label{fig:gap}}{0.95}

\fig{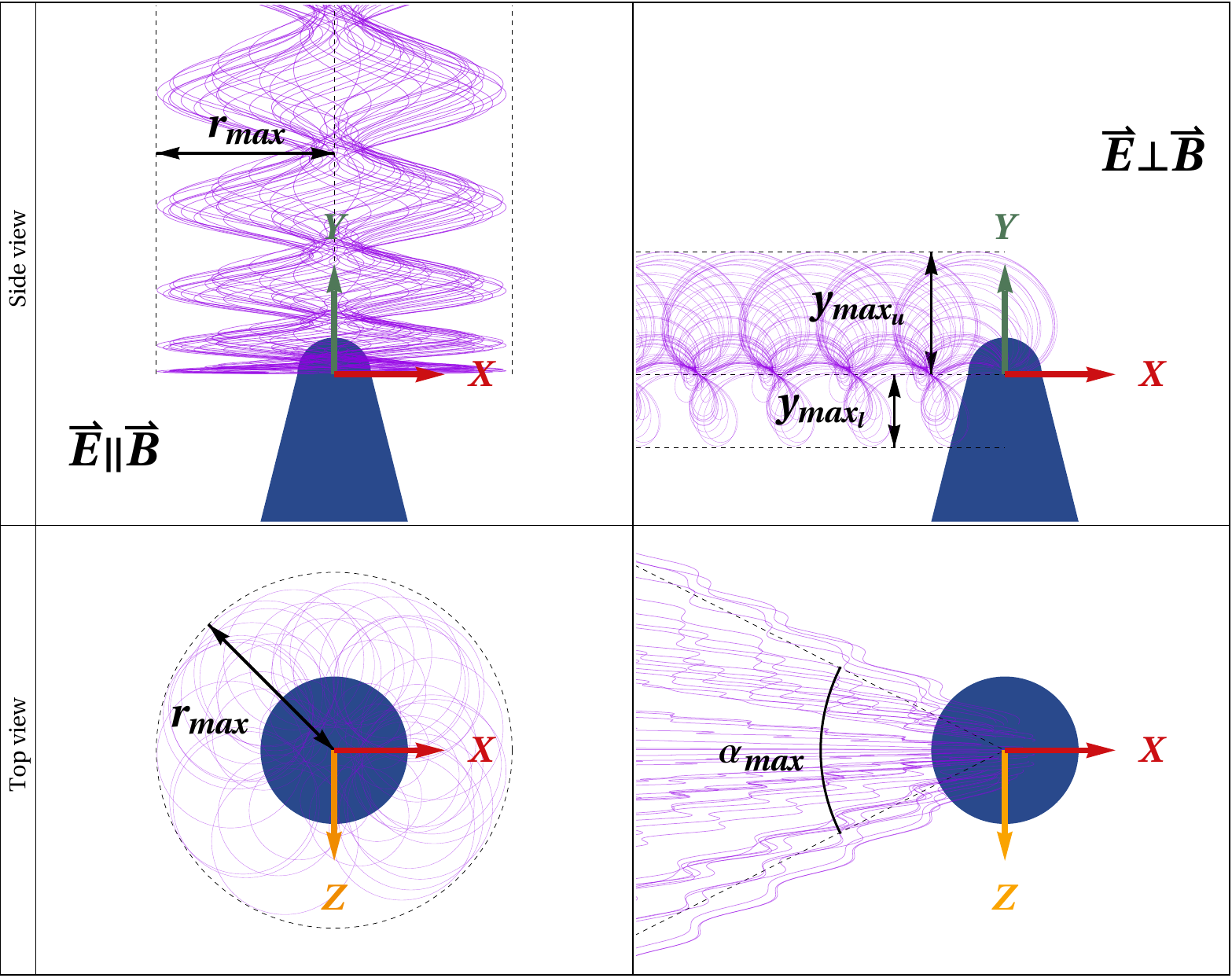}{Top and side view schematic representation of the electron beam dynamics ejected from the cathode tip (blue) for both cases: $\vb{E} \parallel \vb{B}$ and $\vb{E} \bot \vb{B}$, with theoretical parameter illustrations. The purple lines represent the trajectories of individual electrons. The electric field component $E_y$ and the charge-to-mass ratio $q_e/m_e$ are negative, while the magnetic field components $B_y$ (for $\vb{E} \parallel \vb{B}$) and $B_z$ (for $\vb{E} \bot \vb{B}$) are positive. The initial velocity vectors $\vb{v}_{r0}$ exhibit radial symmetry around the center of the tip in both cases. $r_\x{max}$ is the radius of the amplitude of the helix trajectory of the electrons or the maximal deviation of electron trajectory from the axis of the tip for $\vb{E} \parallel \vb{B}$, $y_{\x{max}_\x{u}}$ and $y_{\x{max}_\x{l}}$ define the maximum thickness of the spiral trajectory for for $\vb{E} \bot \vb{B}$ and $\alpha_\x{max}$ is the beam's opening angle for $\vb{E} \bot \vb{B}$.\label{fig:schem}}{0.95}

The motion of the emitted electron beam is studied both theoretically and computationally. Solutions to specific cases can be obtained analytically, while study of more complicated dynamics may require simulation. These two approaches are compared to verify the validity of our results.

\subsection{Theoretical solution of electron beam motion}\label{sec:theory}

The geometry of the system consists of parallel cathode and anode planes separated by a distance $d$ (Fig.~\ref{fig:gap}). A nanoscale tip of height $h$ and radius $r_\x{tip}$ is assumed to be present on the cathode surface (Fig.~\ref{fig:gap}). Near the tip, the electric field is significantly distorted, resulting in localized field enhancement. This field enhancement can lead to vacuum breakdown due to electron emission from the tip surface~\cite{uts67}. The electrons experience radial acceleration near the tip due to presence of the radial component of the field. At the same time, calculating the electrostatic potential near the tip can be challenging~\cite{zub02}. Instead, following Utsumi's ~\cite{uts67} approach, we assume in the analytical solution that the electrons emitted from the tip possess non-zero initial velocities $v_{r0}$. These velocities are sufficiently large that the electrons follow the same parabolic trajectories as they would if field enhancements were explicitly accounted for. 

Assuming that the emitted electron beam is moving in static and uniform electric and magnetic fields, we can solve the motion analytically. The electrons experience acceleration due to the Lorentz force $\vb F = q_e (\vb E + \vb v \times \vb B)$, where $\vb E$ and $\vb B$ are the electric and magnetic fields respectively, $\vb v$ is the velocity and $q_e$ is the charge of the electron. The acceleration of the charged particle depends on the charge-to-mass ratio $q_e/m_e$, so the equations are easily applied to other types of charged particles, such as ions.

We solve these equations and consider two special cases: one where the electric and magnetic fields are parallel $\vb E = \begin{pmatrix} 0 & E_y & 0 \end{pmatrix}$, $\vb B = \begin{pmatrix} 0 & B_y & 0 \end{pmatrix}$ and one where they are perpendicular $\vb E = \begin{pmatrix} 0 & E_y & 0 \end{pmatrix}$, $\vb B = \begin{pmatrix} 0 & 0 & B_z \end{pmatrix}$. These cases are illustrated in Fig.~\ref{fig:schem}.

In the parallel case, we obtain the solution
\begin{equation} \label{eq:parallel}
\begin{cases}
\begin{aligned}
x(t) = x_0 &+ a_m v_{z0} \lb \cos(t/a_m) - 1 \rb \\
&+ a_m v_{x0} \sin(t/a_m), \\
y(t) = y_0 &+ v_{y0} t + \frac{q_e E_y}{2 m_e} t^2, \\
z(t) = z_0 &- a_m v_{x0} \lb \cos(t/a_m) - 1 \rb \\
&+ a_m v_{z0} \sin(t/a_m),
\end{aligned}
\end{cases}
\end{equation}
where $a_m = \frac{m_e}{q_e B_y}$, $\bpm x_0 & y_0 & z_0 \epm$ is the initial position of the particle and $\bpm v_{x0} & v_{y0} & v_{z0} \epm$ is the initial velocity. The constant $a_m$ describes the influence of the magnetic field strength on the electron beam, which affects the amplitude of the beam oscillation. 

\dualfigc
{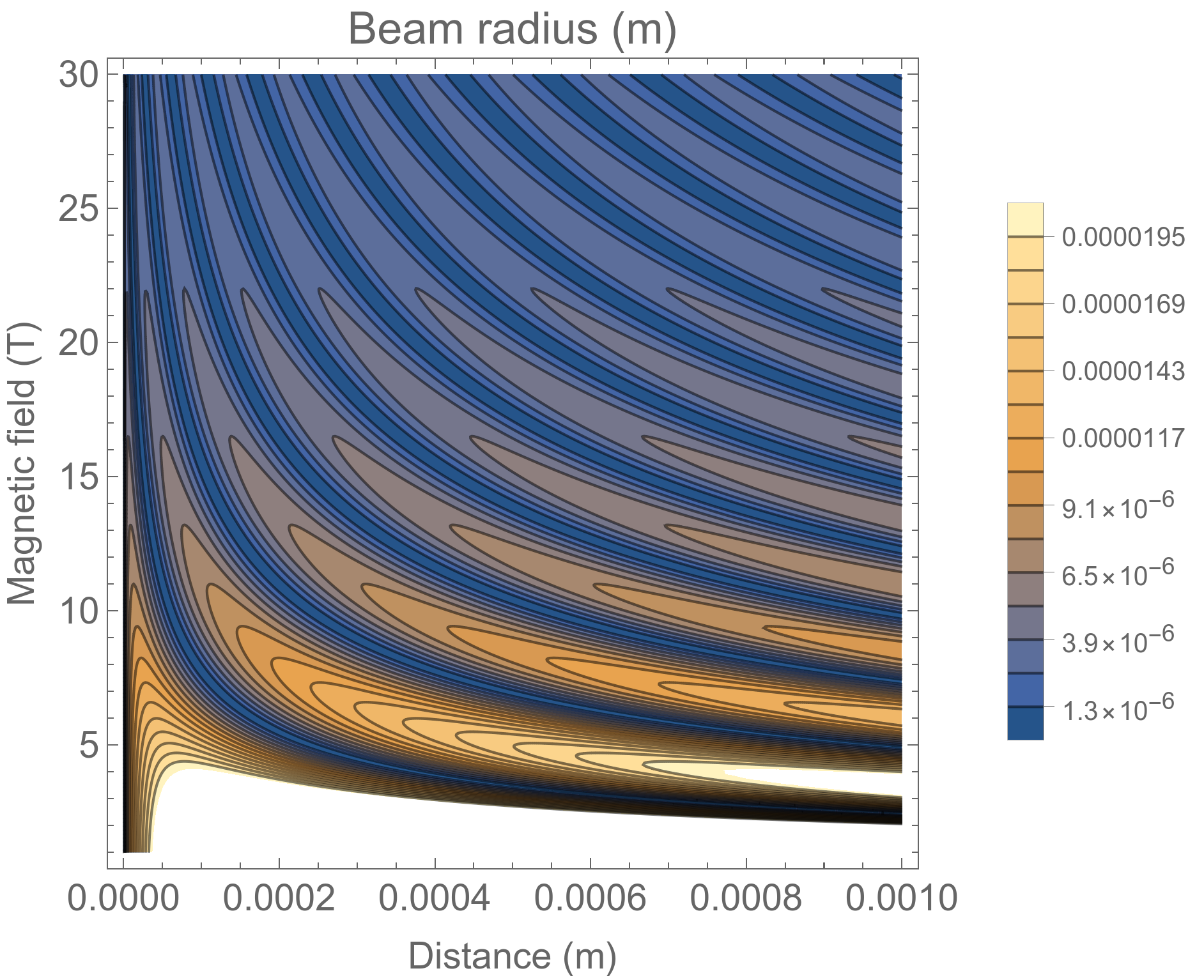}{Beam radius as a function of distance $y$ and magnetic field $B$.\label{fig:beamr}}
{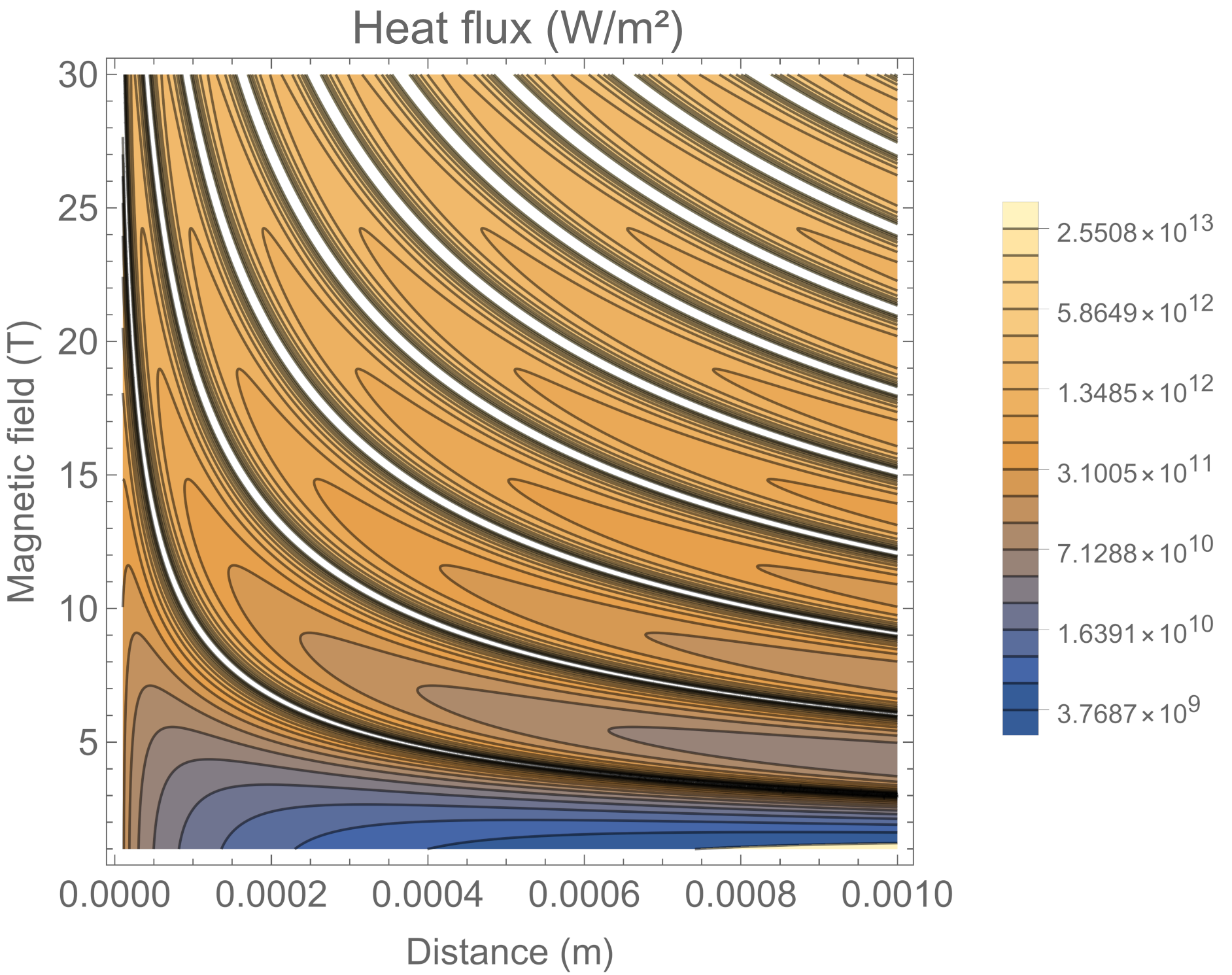}{Heat flux as a function of distance $y$ and magnetic field $B$.\label{fig:beamh}}
{Electron beam quantities in the parallel case ($\vb E \parallel \vb B$) on the anode surface, calculated based on theory by assuming an emitted current of $I_e = 1 \us{mA}$.}

By assuming that the electrons follow a parabolic trajectory, we can get an estimate for the beam radius $r_0(y) = \sqrt{4 \eta h y}$, where $h$ is the height of the emitter and $\eta$ is the spreading factor, which depends on the tip shape. For a hemispherical tip, we assume $\eta = 0.5$~\cite{uts67}.
The ratio of the beam radius $r$ and beam radius without magnetic field $r_0$ can be obtained from theory by calculating the radial velocity
\be\label{eq:v0}
v_{r0} = \frac{\D r_0(y(t))}{\D t} = \sqrt{\frac{2 \eta h q_e}{m_e} E_y},
\ee
and assuming
$
x_0 = y_0 = z_0 = 0, \
v_{y0} = 0.
$
This leads to (e.g., given by Utsumi \cite{uts67})
\begin{align}\label{eq:beam}
\frac{r(y)}{r_0(y)} = \rho \left| \sin\left( \rho^{-1} \right) \right|, \
\rho = \frac{a}{B} \sqrt{\frac{E}{y}},
\end{align}
where $a = \sqrt{\frac{2 m_e}{q_e}} = 3.37 \times 10^{3} \us{nm} \us{T} \usp{V}{-1/2}$, $E$ is electric field, $B$ is magnetic field and $y$ is position. Based on this equation, we expect the radius of the beam to oscillate with a constant amplitude, depending on the magnitudes of the electric and magnetic fields as given by the parameter $\rho$. 

The beam radius depends on distance and magnetic field as shown in Fig. \ref{fig:beamr}, where parameters from Sec.~\ref{sec:results2} are used. We can see that both the amplitude and period of the beam radius depend on the magnetic field $B$. For a larger magnetic field, the beam produced will be narrower and the frequency of oscillation higher. The maximum beam radius can be calculated as $r_\x{max} = \sqrt{r_\x{tip}^{2}+(a_m v_{r0})^{2}}+ \lvert a_m v_{r0}\rvert$ (Fig.~\ref{fig:schem}).

The beam radius can be used to calculate the heat flux at the anode. The power deposited by the electron beam at the anode is $V_0 I_e$, where $V_0$ is the applied voltage at the anode and $I_e$ is the emission current. The heat flux depends on the area of the impacting beam, i.e. the beam radius $r$. The heat flux is
\be\label{eq:heatf}
P = \frac{V_0 I_e}{\pi r^2} = V_0 j = E_0 d j,
\ee
where $j$ is current density, $E_0$ is the applied macroscopic electric field and $d$ is gap distance. We can see that heat flux increases not only with current density, but also with voltage. This shows that if we keep the electric field $E_0$ constant, increasing the gap size will also affect the heat flux. Heat flux at different magnetic fields and distances in shown in Fig.~\ref{fig:beamh} for a beam with a current of $I_e = 1 \us{mA}$. Where the beam radius reaches 0, singularities are present in the heat flux. Although in reality the electron beam always has some finite diameter, this suggests very large heat fluxes can be achieved on the anode surface.

In the perpendicular case, we have the solution
\begin{equation}\label{eq:perpendicular}
\begin{cases}
\begin{aligned}
x(t) = x_0 &+ a_m v_{y0} + \frac{E_y}{B_z} t \\
&- a_m v_{y0} \cos(t/a_m) \\ 
&+ a_m \lb v_{x0} - \frac{E_y}{B_z} \rb \sin(t/a_m), \\
y(t) = y_0 &- a_m v_{x0} + a_m \frac{E_y}{B_z} \\
&+ a_m \lb v_{x0} - \frac{E_y}{B_z} \rb \cos(t/a_m) \\
&+ a_m v_{y0} \sin(t/a_m), \\
z(t) = z_0 &+ v_{z0} t,
\end{aligned}
\end{cases}
\end{equation}
where $a_m = \frac{m_e}{q_e B_z}$, $\bpm x_0 & y_0 & z_0 \epm$ is the initial position of the particle and $\bpm v_{x0} & v_{y0} & v_{z0} \epm$ is the initial velocity. The electron beam has rotational motion in the $xy$ plane, while in the $z$ direction the motion is not influenced by the electric or magnetic fields. The propagation of particles in $y$ direction will be be limited by $y_\text{max} = a_m b \lb 1 \pm \sqrt{1 + \lb \frac{v_{y0}}{b} \rb^2} \rb + y_0$, where $b = \frac{E_y}{B_z}-v_{r0}$. There is a dependency on the ratio of the electric and magnetic fields, expressed as $E_y / B_z$. This ratio influences the beam dynamics, causing the beam to expand outwards with an opening angle $\alpha_\text{max}$, which is defined as: $\alpha_\text{max} = \pi - 2 \arctan \left\lvert\frac{E_y}{{B_z}v_{r0}}\right\rvert$ (Fig.~\ref{fig:schem}).

As the ratio $\frac{E_y}{{B_z}v_{r0}}$ decreases, the angle distribution of the particles shows a distinct pattern of shifting density. Angles closer to $0\degs$, which coincide with the $x$ direction, become sparsely distributed, whereas those near $\pm 90\degs$ exhibit a higher concentration. In the limiting case, where $\frac{E_y}{{B_z}v_{r0}} \rightarrow 0, \ \alpha=\pi$ and $v_x = 0$. Consequently, the beam propagates only in the $\pm z$ directions, and then the radius of the beam is given by $r=2a_m v_{r0}+r_\x{tip}$.

\subsection{Simulation methods}\label{sec:simulation}

Simulation of electron beams under electric and magnetic fields is well suited for the particle-in-cell method (PIC)~\cite{tsk07}, which we use to calculate trajectories of emitted electrons while including electron-electron interactions using Monte Carlo collisions (MCC)~\cite{tak77}. We use the FEMOCS code~\cite{ves20} to perform these simulations in 3D, with electron emission calculations performed using GETELEC~\cite{kyr17}. Electric field, current and heat calculations are performed using the finite element method using the deal.II library~\cite{dea21}, with meshes generated using Gmsh~\cite{geu09} and TetGen~\cite{han15}. The electric field with space charge is solved from Poisson's equation. The current and heat distributions in the cathode are solved from the continuity and heat equations respectively. A more detailed description of our current simulation model is given in~\cite{koi24}.

On the cathode surface (Cu), a nanotip is assumed to be present, causing local field enhancement and emission of electrons. These electrons travel across the cathode-anode gap and impact the anode. PIC simulation was performed for electrons with a time step of $\Delta t_\x{PIC} = 0.1 \us{fs}$ and superparticle weight of $w_\x{SP} = 10$. In the electron beam, we perform elastic and Coulomb collisions between electrons.

We can calculate the distribution of electrons impacting the anode (Cu) and determine the current density on the anode surface to get the temperature distribution due electron bombardment~\cite{zho19}. The current density is used for finite element (FEM) heating calculations, performed using COMSOL (version 6.2). In this model, we calculate the heat rate in the anode as $P = V_0 j / z_d$, where $j$ is current density, $V_0$ is the applied voltage and $z_d$ is the electron beam CSDA (continuous-slowing-down approximation) range in Cu. When $z > z_d, P = 0$. The base of the anode is assumed to be at a temperature of $T_a = 300$~K. The simulation model includes solid heat transfer inside the anode, electric current calculations, electromagnetic heating (resistive heating) and radiative cooling of the anode surface.

\section{Results}

We performed two types of simulations: PIC simulations to study the influence of the direction and magnitude of the magnetic field on the dynamics of the emitted electron beam and FEM heating simulations to calculate the heating of the anode surface due to electron bombardment.

\subsection{Direction of magnetic field}

\trifigw
{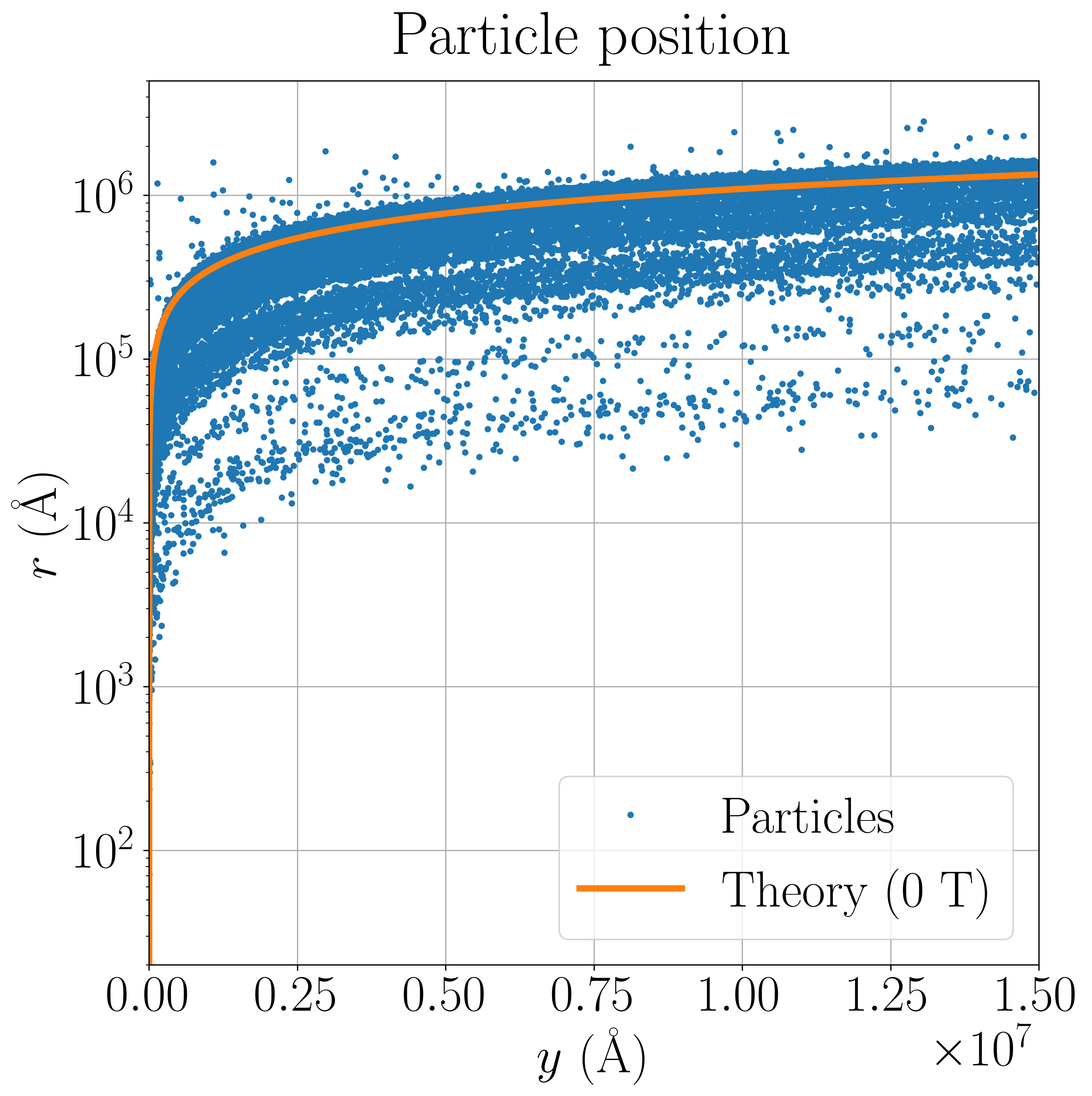}{$B_y = 0$~T.\label{fig:beamy0}}
{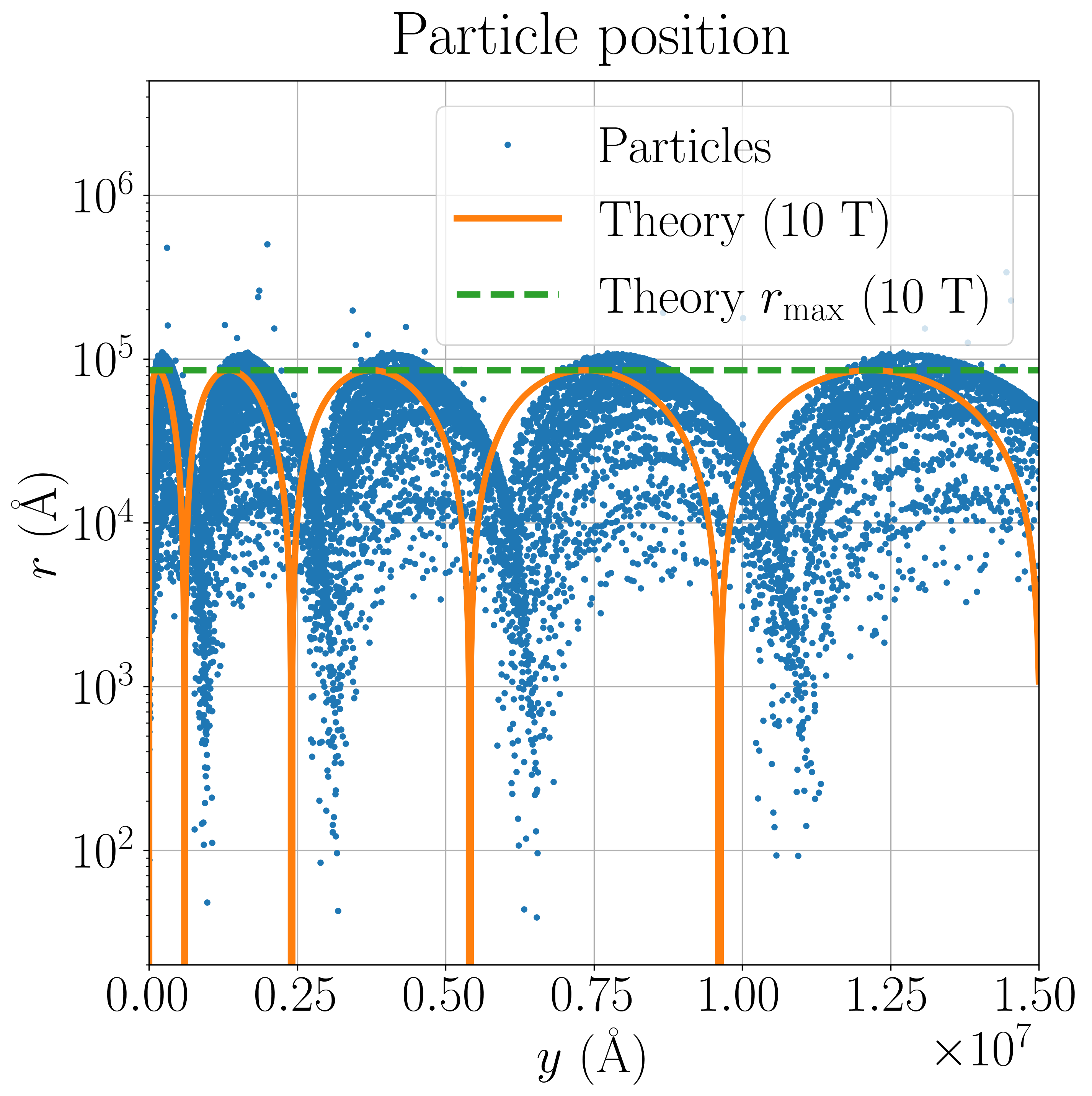}{$B_y = 10$~T.\label{fig:beamy10}}
{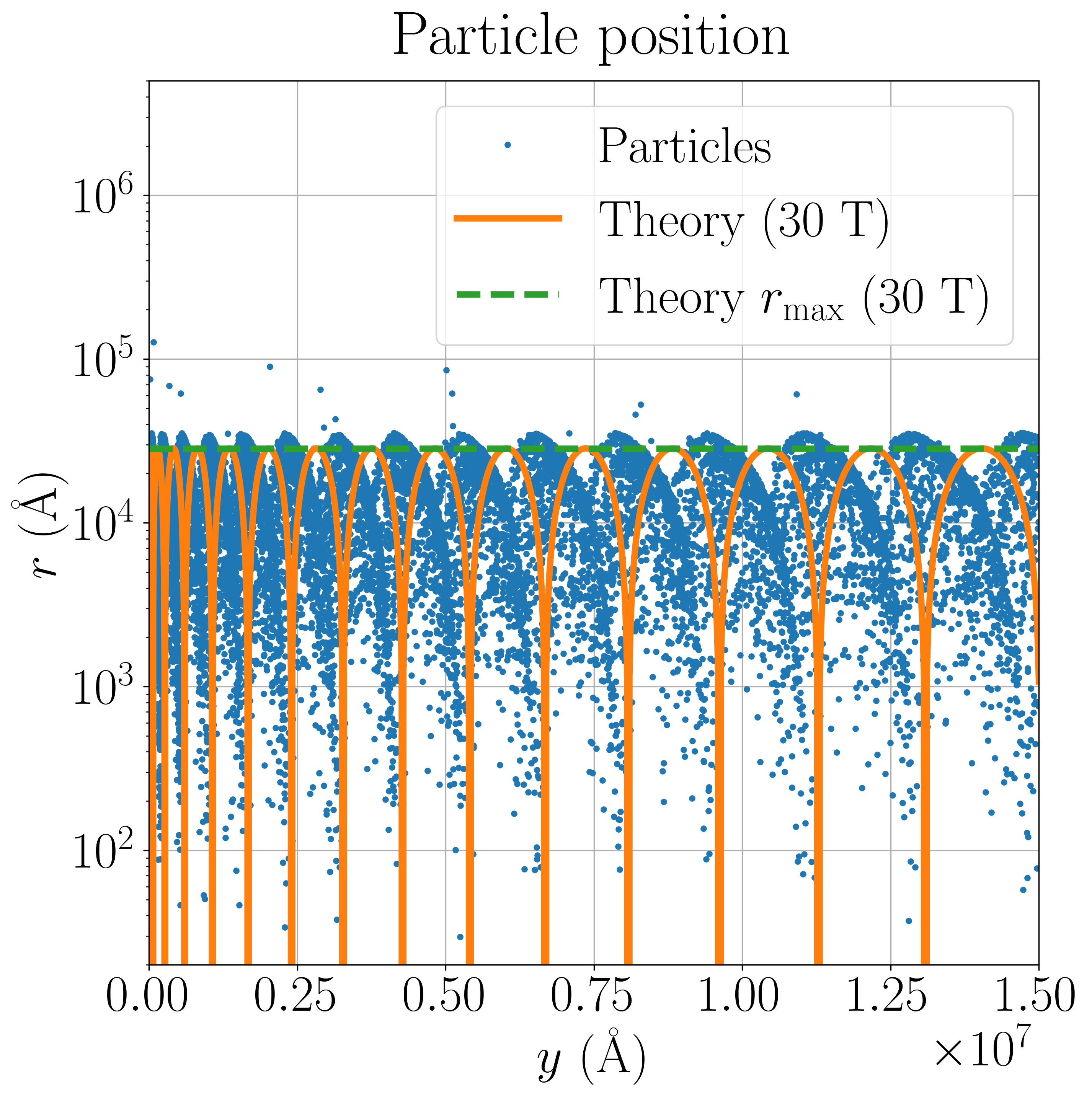}{$B_y = 30$~T.\label{fig:beamy30}}
{Modulation of the electron beam shape at different magnetic fields in the case of parallel electric and magnetic fields ($\vb{E} \parallel \vb{B}$). In the legend, ``Theory'' refers to the results of Eq.~\ref{eq:parallel} and $r_\x{max}$ is the radius of maximal radial deviation from the tip axis for the $\vb{E} \parallel \vb{B}$ geometry (Fig.~\ref{fig:schem}).}
\begin{figure*}
  \centering
  \begin{subfigure}[t]{.32\textwidth}
  \centering
    \includegraphics[width=1.0\linewidth]{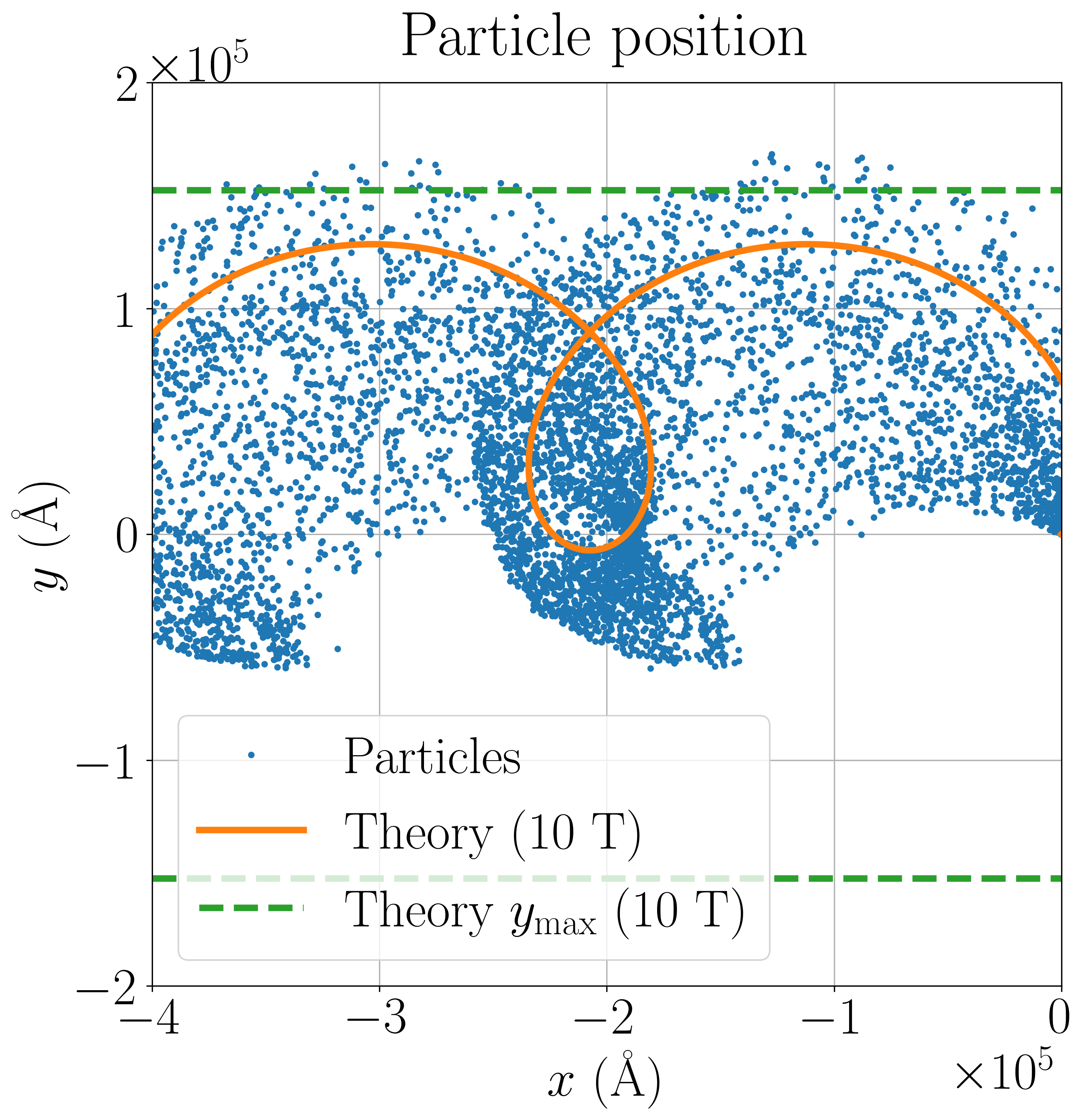}
    \caption{$B_z = 10$~T (side).\label{fig:beamxy10}}
  \end{subfigure}
  \begin{subfigure}[t]{.32\textwidth}
    \centering
    \includegraphics[width=1.0\linewidth]{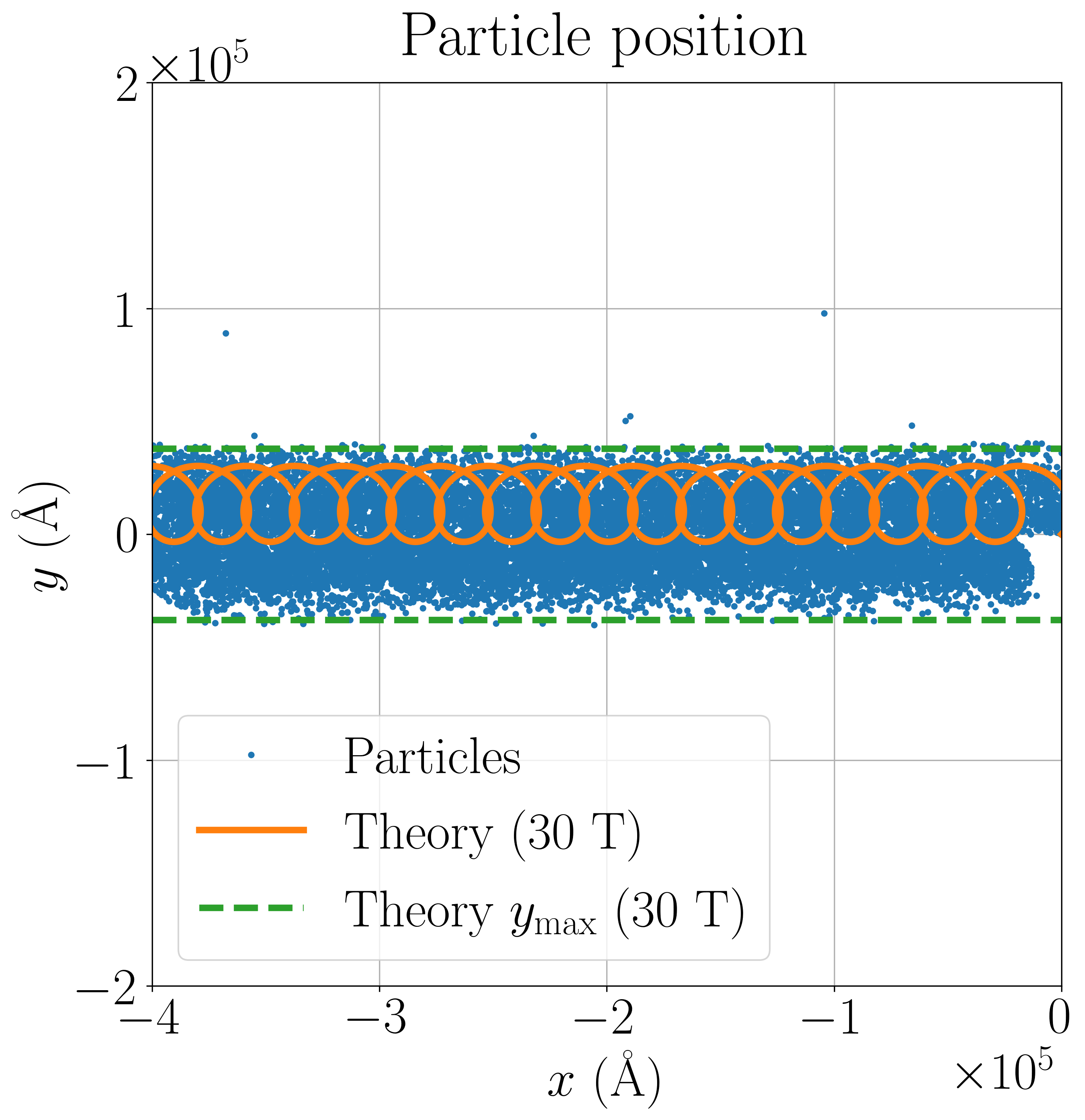}
    \caption{$B_z = 30$~T (side).\label{fig:beamxy30}}
  \end{subfigure}
  \begin{subfigure}[t]{.32\textwidth}
    \centering
    \includegraphics[width=1.0\linewidth]{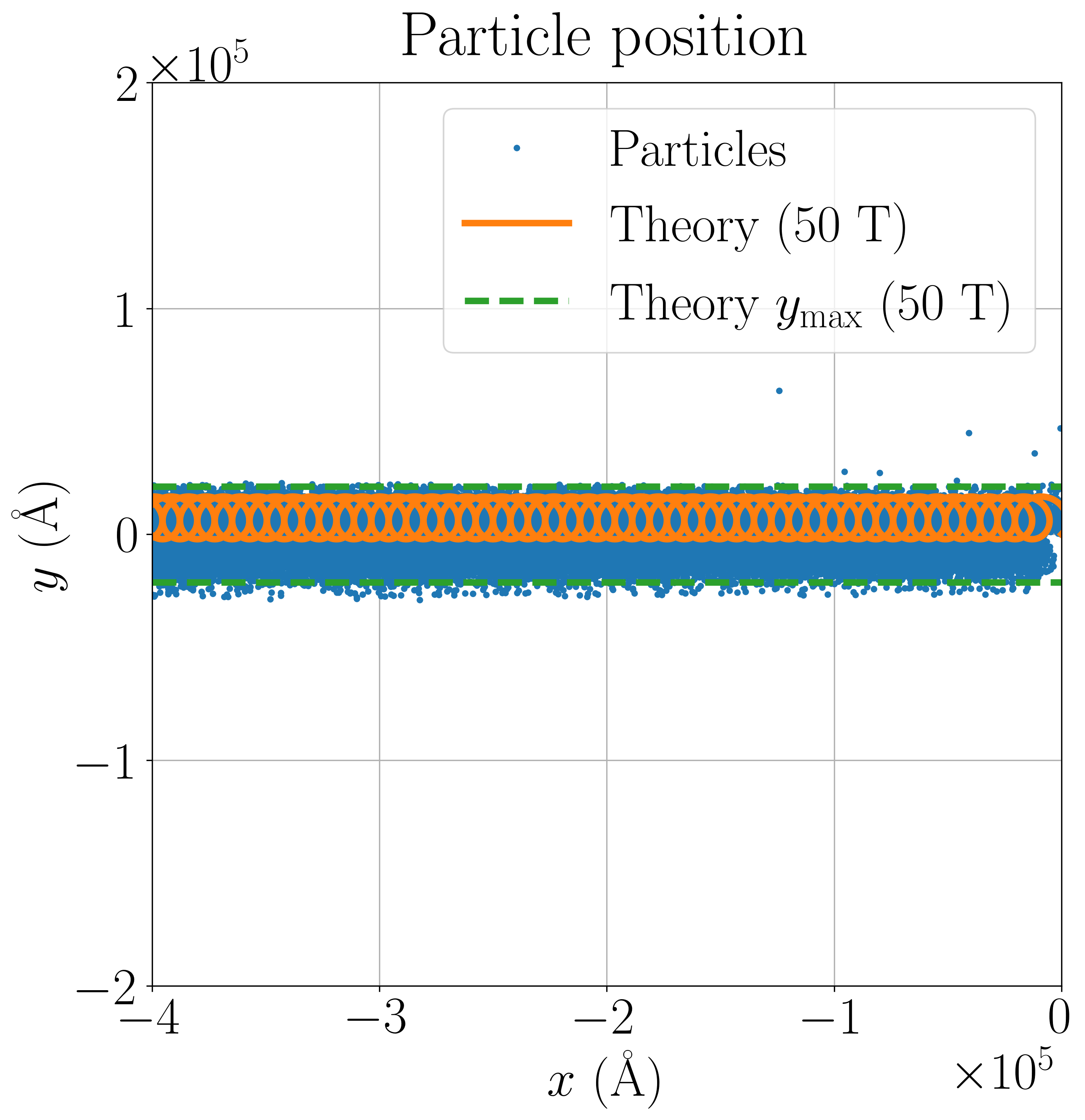}
    \caption{$B_z = 50$~T (side).\label{fig:beamxy50}}
  \end{subfigure}
  \begin{subfigure}[t]{.32\textwidth}
  \centering
    \includegraphics[width=1.0\linewidth]{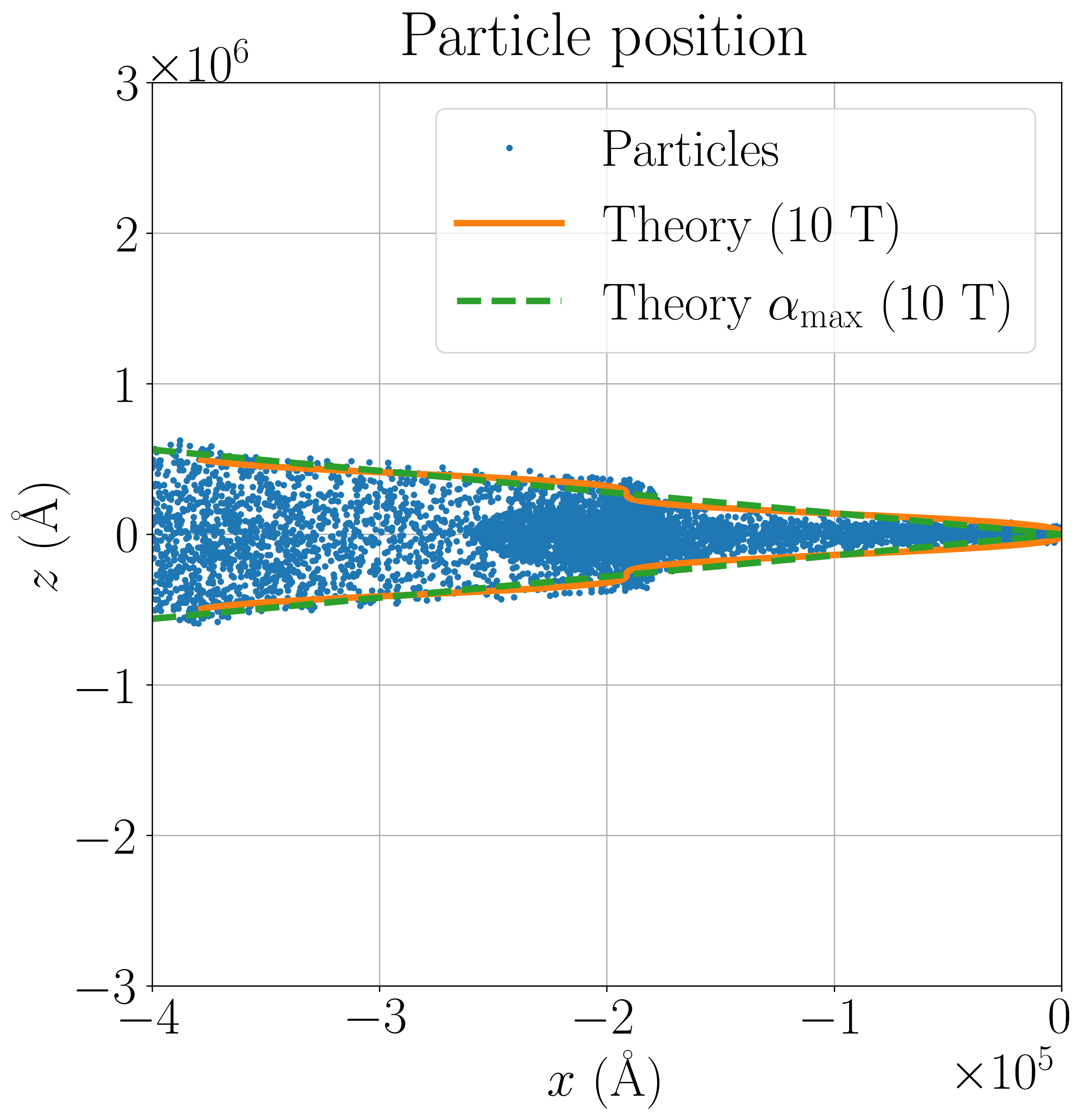}
    \caption{$B_z = 10$~T (top).\label{fig:beamxz10}}
  \end{subfigure}
  \begin{subfigure}[t]{.32\textwidth}
    \centering
    \includegraphics[width=1.0\linewidth]{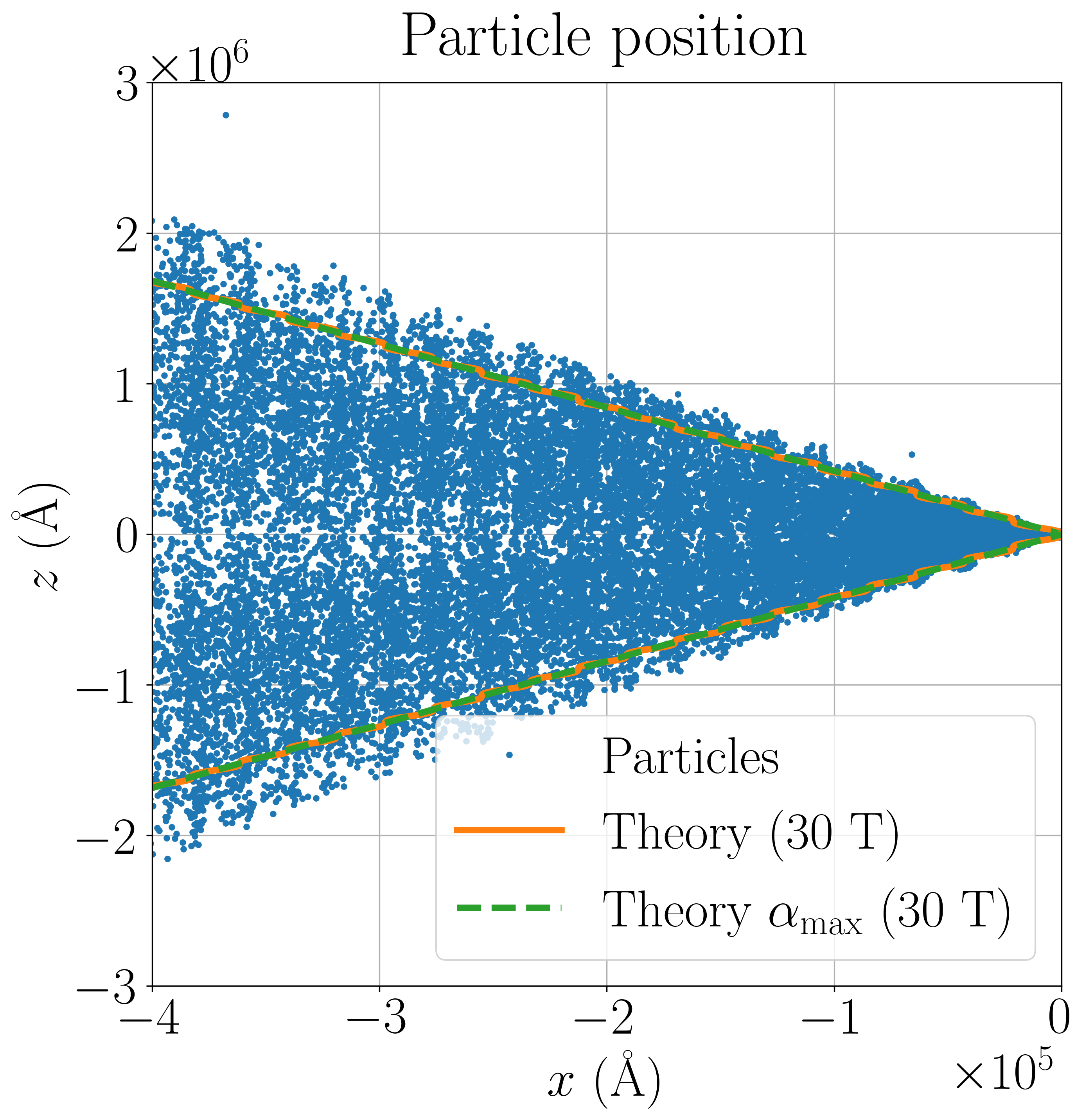}
    \caption{$B_z = 30$~T (top).\label{fig:beamxz30}}
  \end{subfigure}
  \begin{subfigure}[t]{.32\textwidth}
    \centering
    \includegraphics[width=1.0\linewidth]{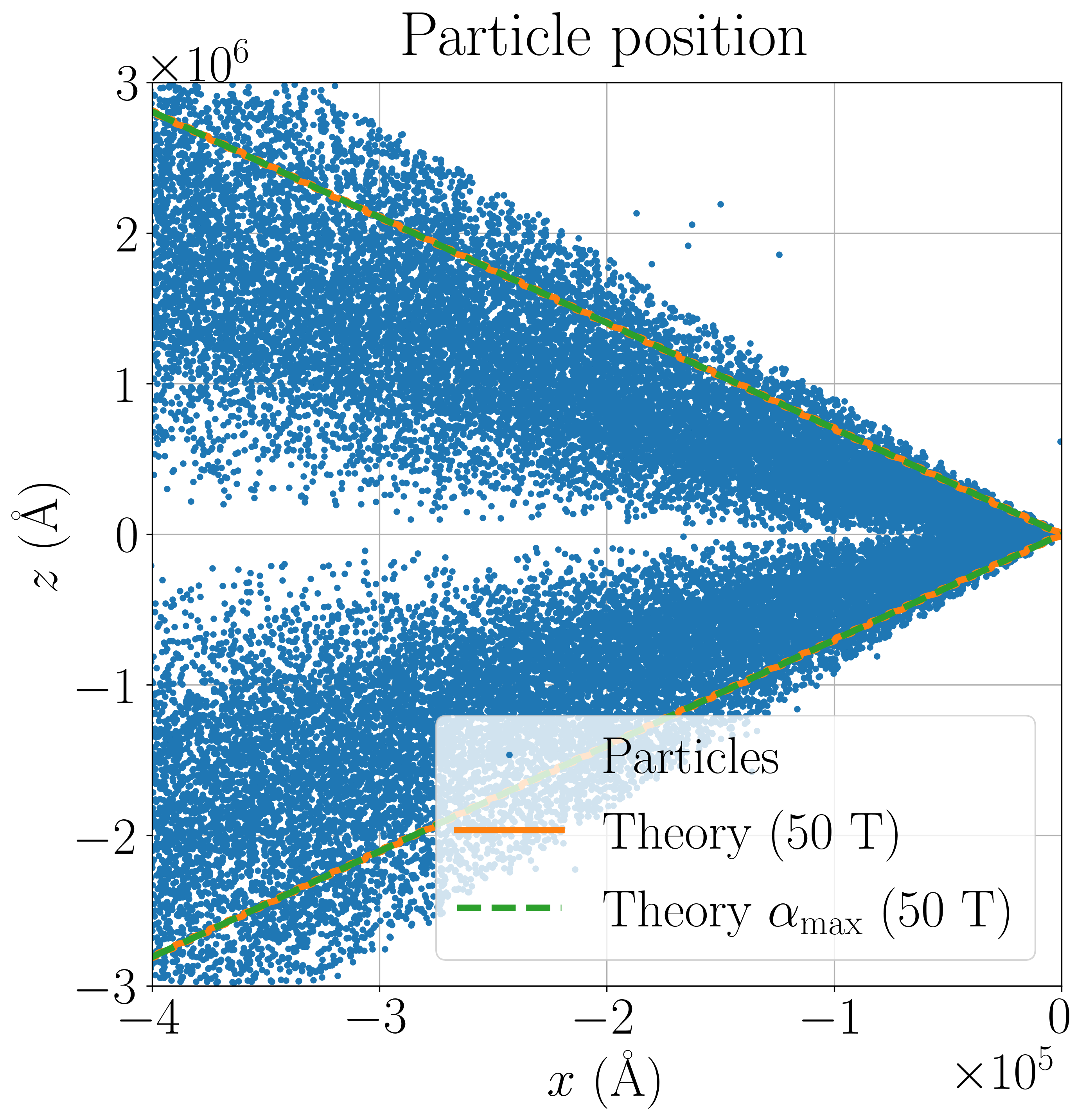}
    \caption{$B_z = 50$~T (top).\label{fig:beamxz50}}
  \end{subfigure}
  \caption{Modulation of the electron beam shape at different magnetic fields in the case of perpendicular electric and magnetic fields ($\vb{E} \bot \vb{B}$). The top and bottom rows show the side and top views of the beam shapes for the corresponding values of the magnetic fields. In the legend, ``Theory'' refers to the results of Eq.~\ref{eq:perpendicular} and $y_\x{max}$ is the maximal vertical deviation of the electron trajectory for the $\vb{E} \bot \vb{B}$ geometry (Fig.~\ref{fig:schem}).}
\end{figure*}

We simulated two cases: one where the magnetic field is parallel with the electric field (Fig.~\ref{fig:beamy0}--\ref{fig:beamy30}), and one where it is perpendicular to it (Fig.~\ref{fig:beamxy10}--\ref{fig:beamxz50}). Depending on the direction of the magnetic field, the electron beam can be either focused or defocused. In both parallel and perpendicular cases, the simulated electron beam can be seen to produce a shape similar to our theoretical solutions (Eq.~\ref{eq:parallel} and \ref{eq:perpendicular}). This verifies that the simulation results roughly follow the theoretical solution while showing that the beam dynamics is not fully captured by theory alone.

In the parallel case (Fig.~\ref{fig:beamy0}--\ref{fig:beamy30}), we see that the electrons experience inward radial acceleration and are focused to a point. The electrons are accelerated by the electric field from the cathode surface towards the anode. The magnetic field causes the electron beam to rotate around the $y$ axis, which leads to forces perpendicular to the axis. The electron beam radius oscillates with a constant amplitude and increasing period. These results show how the application of a magnetic field causes the electron beam to impact the anode surface with a significantly smaller radius compared to the case with no magnetic field.

In the perpendicular case (Fig.~\ref{fig:beamxy10}--\ref{fig:beamxz50}), the electron beam expands in the direction of the magnetic field and rotates around the magnetic field axis. The beam propagates in a direction perpendicular to the emitting tip and does not impact the anode. The amplitude and frequency of rotation depend on the magnitude of the applied magnetic field, while the opening angle of the beam depends on the ratio of the electric and magnetic fields as calculated in Sec.~\ref{sec:theory}. For the theoretical solutions, the path of the electrons depends on their initial velocities, which is assumed to be given by Eq.~\ref{eq:v0}. The initial velocity is assumed to be evenly divided between the $x$ and $y$ components in the $xy$ plots, while in the $xz$ plots it is in the $z$ component. In the simulations, the initial velocity depends on the exact geometry, which results in differences observed in the results. These results show that the direction of the magnetic field can significantly influence the beam, which can change the magnitude of heating on the anode side.

\subsection{Heating due to magnetic field}\label{sec:results2}

\trifighs
{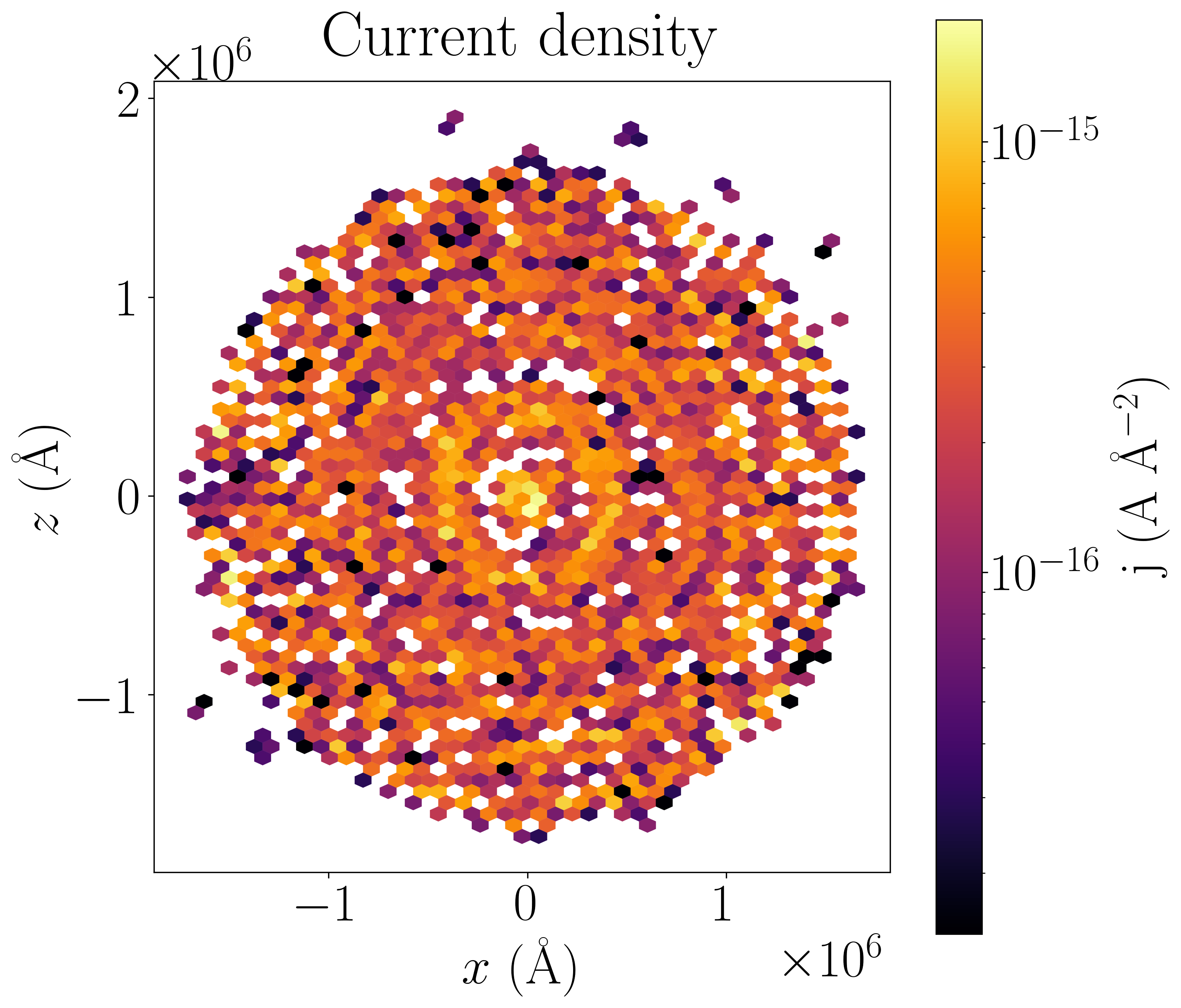}{$B_y = 0$~T.\label{fig:cdens0}}
{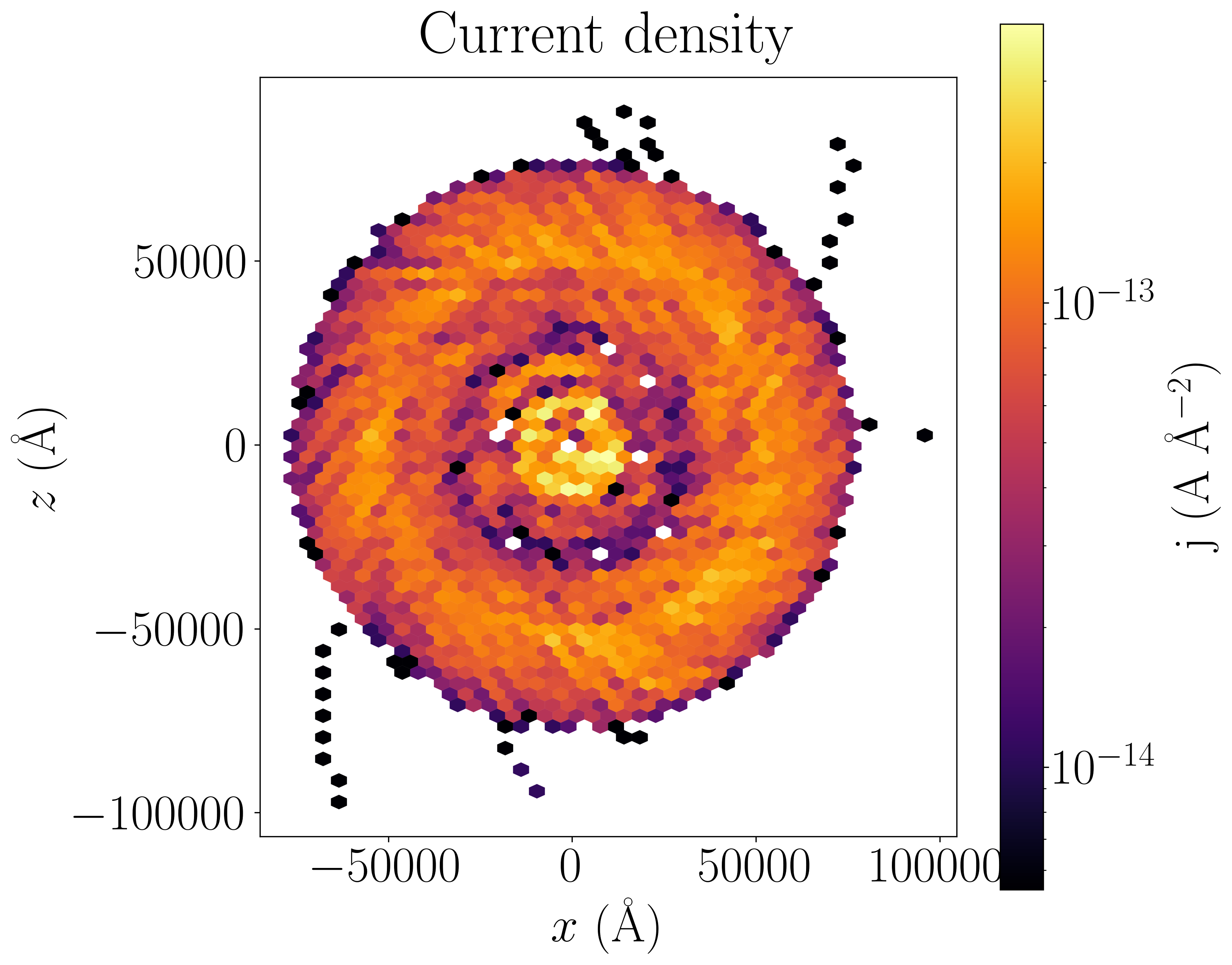}{$B_y = 10$~T.\label{fig:cdens10}}
{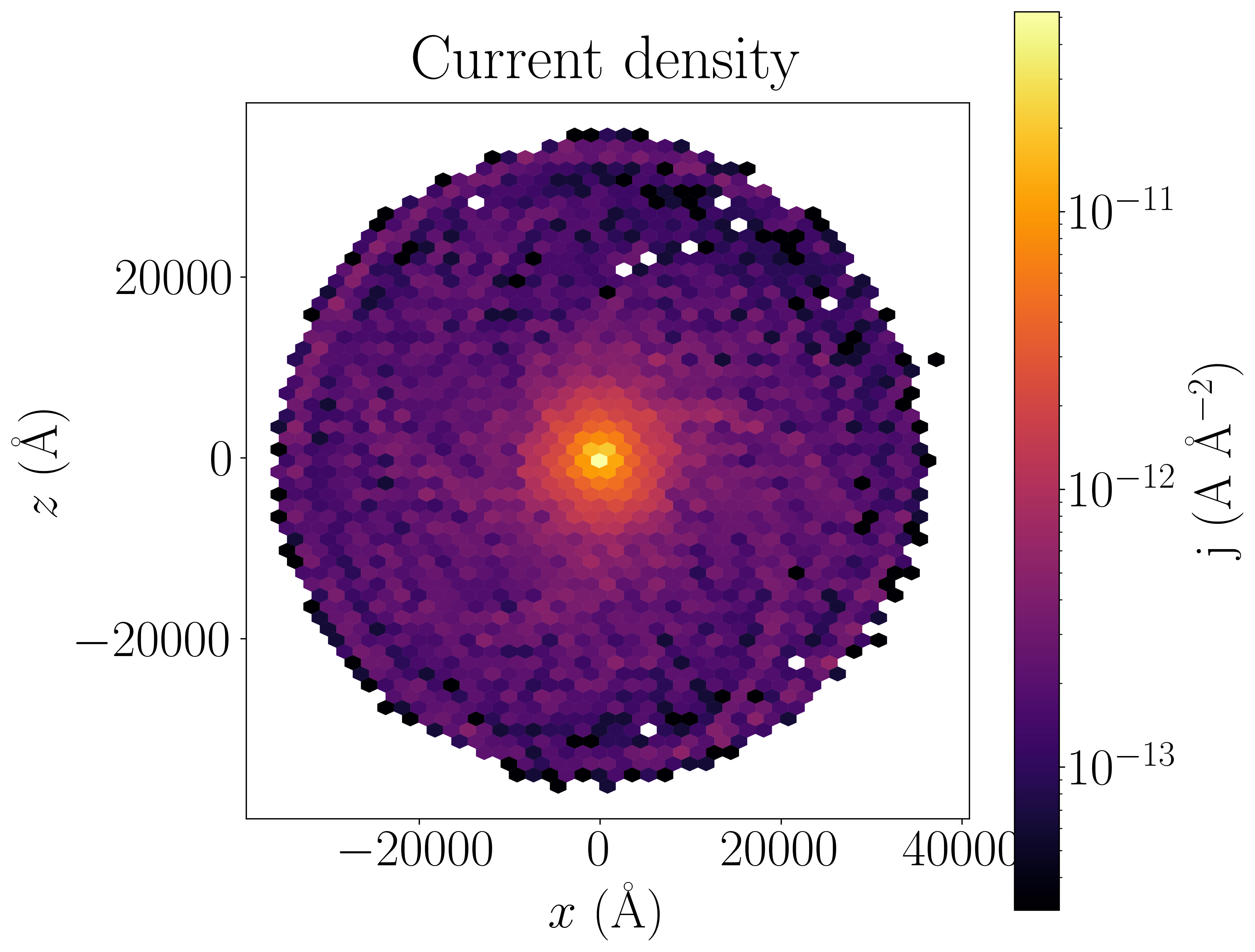}{$B_y = 30$~T.\label{fig:cdens30}}
{Spatial electron current density on anode surface in the parallel case ($\vb{E} \parallel \vb{B}$) for the different magnitudes of the magnetic field, obtained by PIC simulation. The color of each hexagonal pixel represents the 2.5 ps time-averaged electron current density on the anode surface. Increasing the magnetic field from 0~T to 10~T reduces the spot diameter to about $1/20$ of its size, reducing the area to just $1/400$ of its original size. Increasing the magnetic field from 10~T to 30~T reduces the diameter further to about $1/2$, or to $1/4$ of the previous area.}{0.82}

\trifigw
{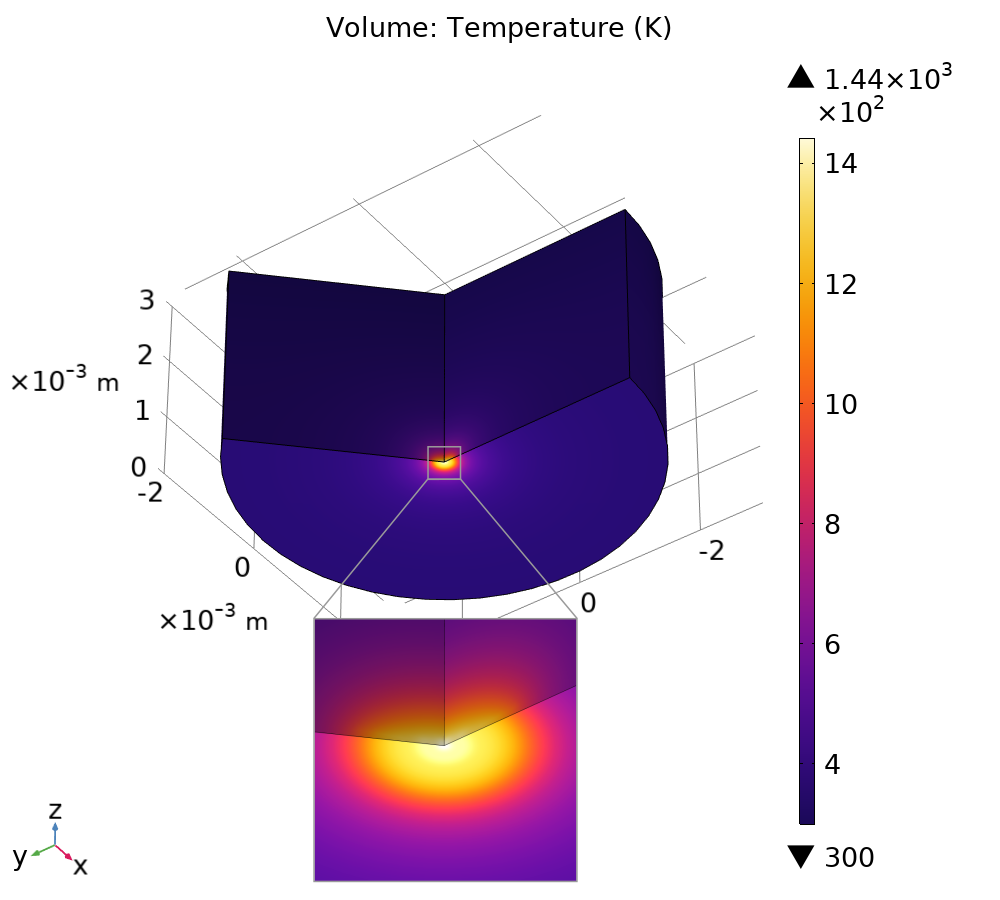}{$B_y = 0$~T.\label{fig:temp0}}
{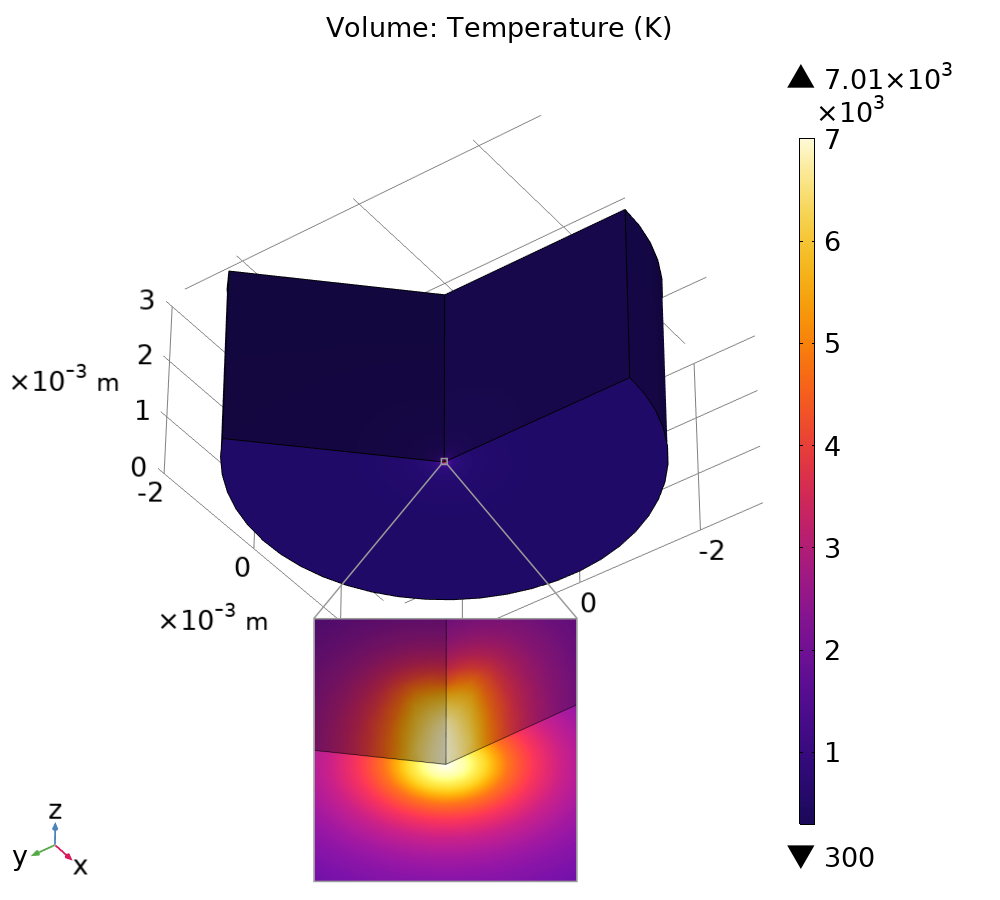}{$B_y = 10$~T.\label{fig:temp10}}
{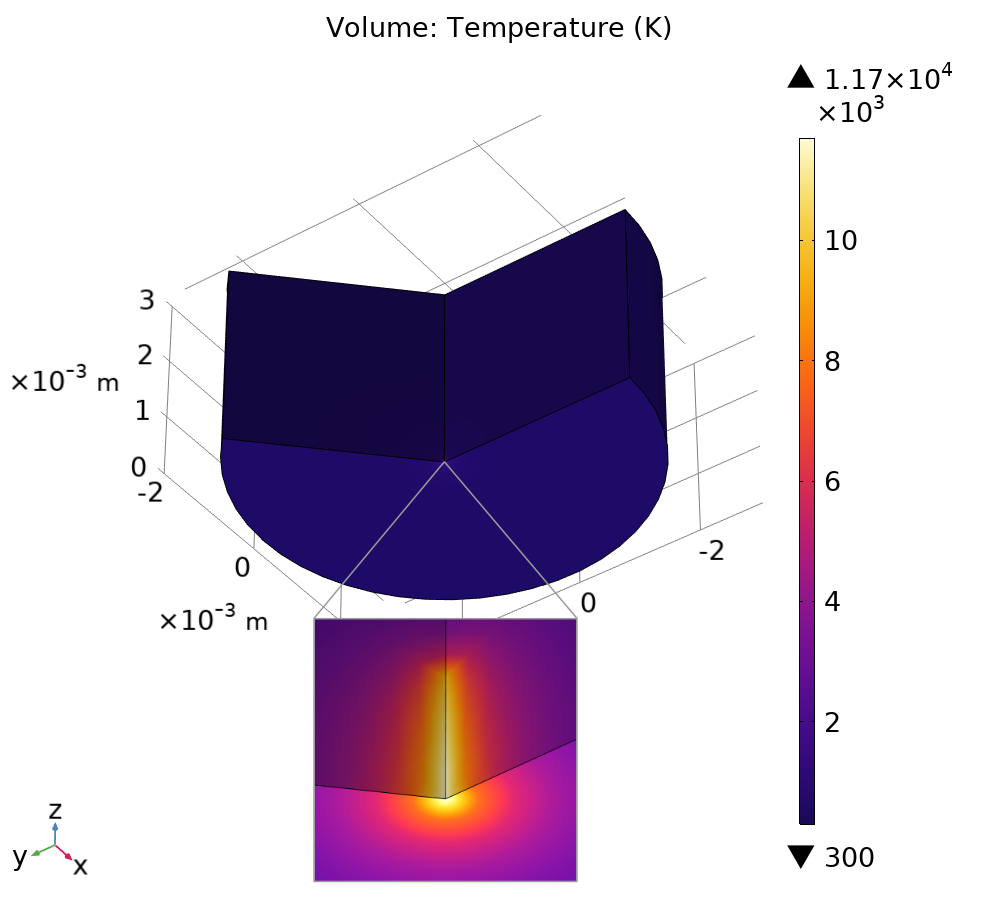}{$B_y = 30$~T.\label{fig:temp30}}
{Spatial plot of anode steady state temperature in the parallel alignment of electric and magnetic fields ($\vb{E} \parallel \vb{B}$) for different values of the magnetic field, obtained by COMSOL (based on the current densities from Fig.~\ref{fig:cdens0}--\ref{fig:cdens30}). The color represents the anode temperature. The maximum temperatures observed were 1440 K at 0 T, 7010 K at 10 T and 11700 K at 30 T.}

Electron beam simulations are performed using PIC with a parallel magnetic field at 0~T, 10~T and 30~T to determine how the magnetic field strength influences heating on the anode side. The distance between the cathode and anode (gap size) is $d = 1.5 \us{mm}$, while the diameter of the cathode and anode is about $300 \us{\textmu m}$. On the cathode surface, there is a tip with a height of $6 \us{\textmu m}$ and apex radius of $15 \us{nm}$. The applied voltage is $V_0 = 80 \us{kV}$, giving a macroscopic field of 53~MV/m and a local field of $E_0 = 10 \us{GV}/\un{m}$ at the tip apex. This results in an electron emission current of about $I_e = 2.5 \us{mA}$.

The simulated electron beam shape is plotted with the theoretical prediction (Eq.~\ref{eq:beam}) in Fig.~\ref{fig:beamy0}--\ref{fig:beamy30}. We can see that the general agreement is good, with some variation in the electron paths depending on where they were emitted on the cathode surface and their initial velocities. There may be some influence due to Coulomb repulsion of the electrons, which could defocus the beam. In the theoretical predictions, the spreading factor is assumed to be $\eta = 0.5$, which can also lead to differences compared to the simulation results. In the 0~T case, the electrons can be seen following a parabolic trajectory, as expected. In the 10~T and 30~T cases the electron beam is confined by the magnetic field so that its diameter stays within constant limits.

The current density of the impacting electrons on the anode surface is calculated based on the PIC simulations (Fig.~\ref{fig:cdens0}--\ref{fig:cdens30}). This is done by calculating the density and velocity of electrons above the anode surface ($j = \rho v$) and time-averaging over 100 frames. It can be seen that increasing the magnetic field from 0~T to 10~T reduces the spot diameter to about $1/20$ of its size, reducing the area to just $1/400$ of its original size. Increasing the magnetic field from 10~T to 30~T reduces the diameter further to about $1/2$, or to $1/4$ of the previous area. As the emitted current stays roughly constant, this reduction in area results in a corresponding increase in current density. The rotational motion of the electron beam in the magnetic field can be seen in the current density distributions as spiraling streaks.

Based on the calculated current densities (Fig.~\ref{fig:cdens0}--\ref{fig:cdens30}), we calculate the steady-state temperature distribution in the anode using COMSOL as described in Sec.~\ref{sec:simulation}. This is done for the 0~T (Fig.~\ref{fig:temp0}), 10~T (Fig.~\ref{fig:temp10}) and 30~T (Fig.~\ref{fig:temp30}) cases. According to the NIST ESTAR database~\cite{ber99}, the CSDA range for electrons at 80~keV is $1.535 \tp{-2} \us{g} \usp{cm}{-2}$ or $z_d = 17.1 \us{\textmu m}$ for Cu, which we use as the penetration depth of the beam. The maximum temperatures observed were 1440~K at 0~T, 7010~K at 10~T and 11700~K at 30~T. Increasing the magnetic field from 0~T to 10~T results in a temperature increase of over 5000~K, from the melting point to well above the boiling point. This would result in the evaporation of neutral atoms from the surface, which could start plasma formation. A further increase to 30~T leads to even more extreme temperatures, demonstrating how the magnetic field can lead to significant heating. However, these temperatures are localized to a very small point and would likely rapidly diffuse away unless continuous heating is applied, as instabilities could lead to shifting of the beam. Furthermore, at such temperatures, a phase change would occur, limiting the temperature.

The effect of pulsed heating on the maximum temperature is studied in Fig.~\ref{fig:tpulse}. A time-dependent simulation is run for the cases in Figs.~\ref{fig:temp0}--\ref{fig:temp30} up to $100\us{ns}$. The maximum temperature reached at time $t$ in Fig.~\ref{fig:tpulse} is the instantaneous local temperature resulting from a pulse of width $t$. For example, a rectangular wave with $1\us{MHz}$ frequency and $10^{-3}$ duty cycle would have a pulse width of $1\us{ns}$. A $1\us{ns}$ pulse resulted in a maximum temperature of 300~K at 0~T, 333~K at 10~T and 959~K at 30~T, showing a major reduction in heating of the anode surface.

\fig{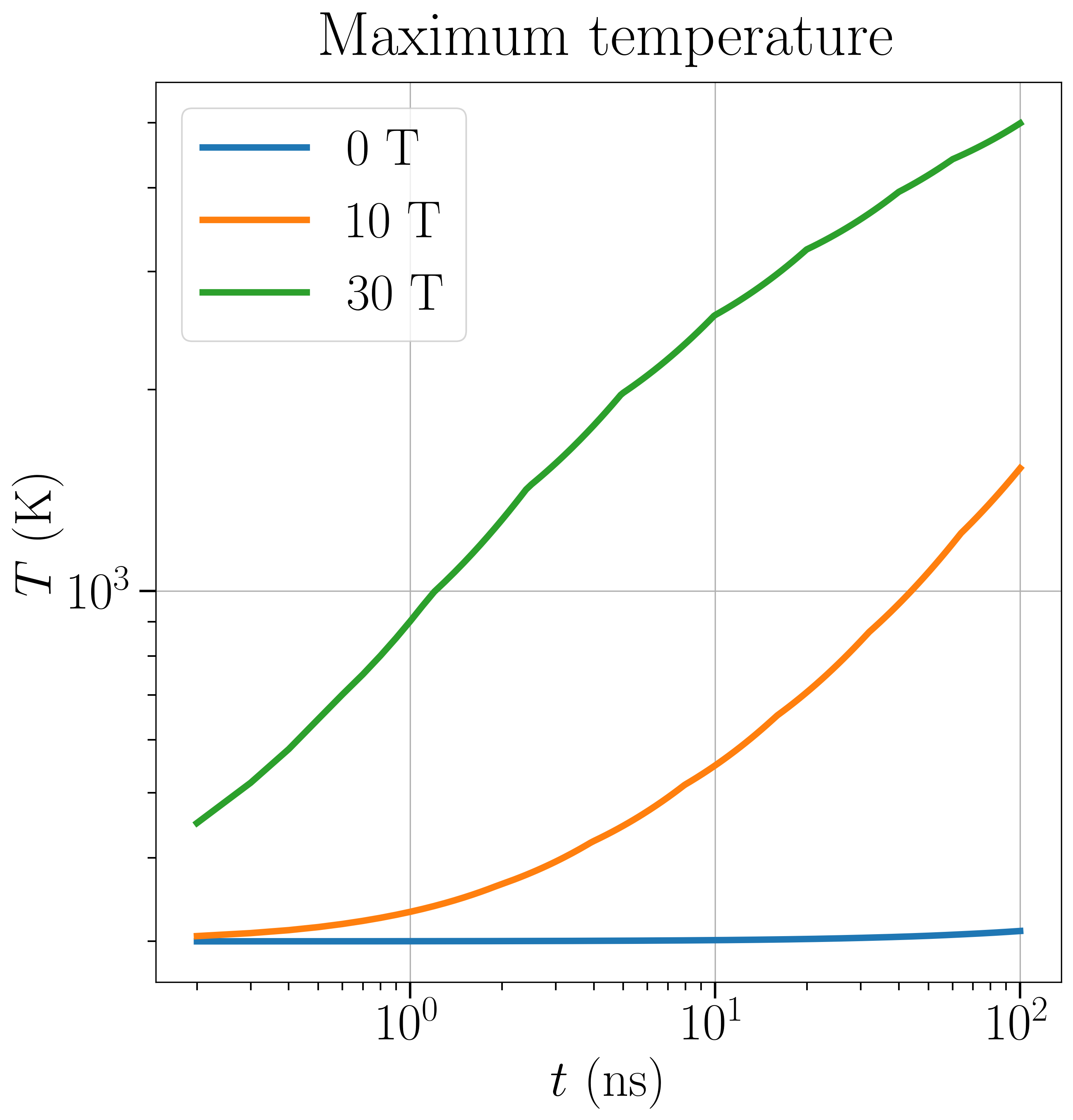}
 {Time-dependent maximum temperature on anode surface (produced by a pulse of width $t$) for the 0~T (Fig.~\ref{fig:temp0}), 10~T (Fig.~\ref{fig:temp10}) and 30~T (Fig.~\ref{fig:temp30}) cases.\label{fig:tpulse}}{0.7}

The heating of the anode surface is additionally studied by varying the spot radius $r_s$ and power density $Q_s$ and calculating the steady-state temperature on the anode surface. The heat flux within the spot is assumed to be uniformly distributed, and the maximum temperature on the anode surface is evaluated. A smaller spot radius and higher power density are the factors that contribute to greater heating of the anode surface. The results of these simulations are shown in Fig.~\ref{fig:tanode}, which agree with theoretical and experimental values for the critical anode power density in Cu~\cite{uts67}. The theoretical prediction for Cu in~\cite{uts67} is given as a red line for reference.

\fig{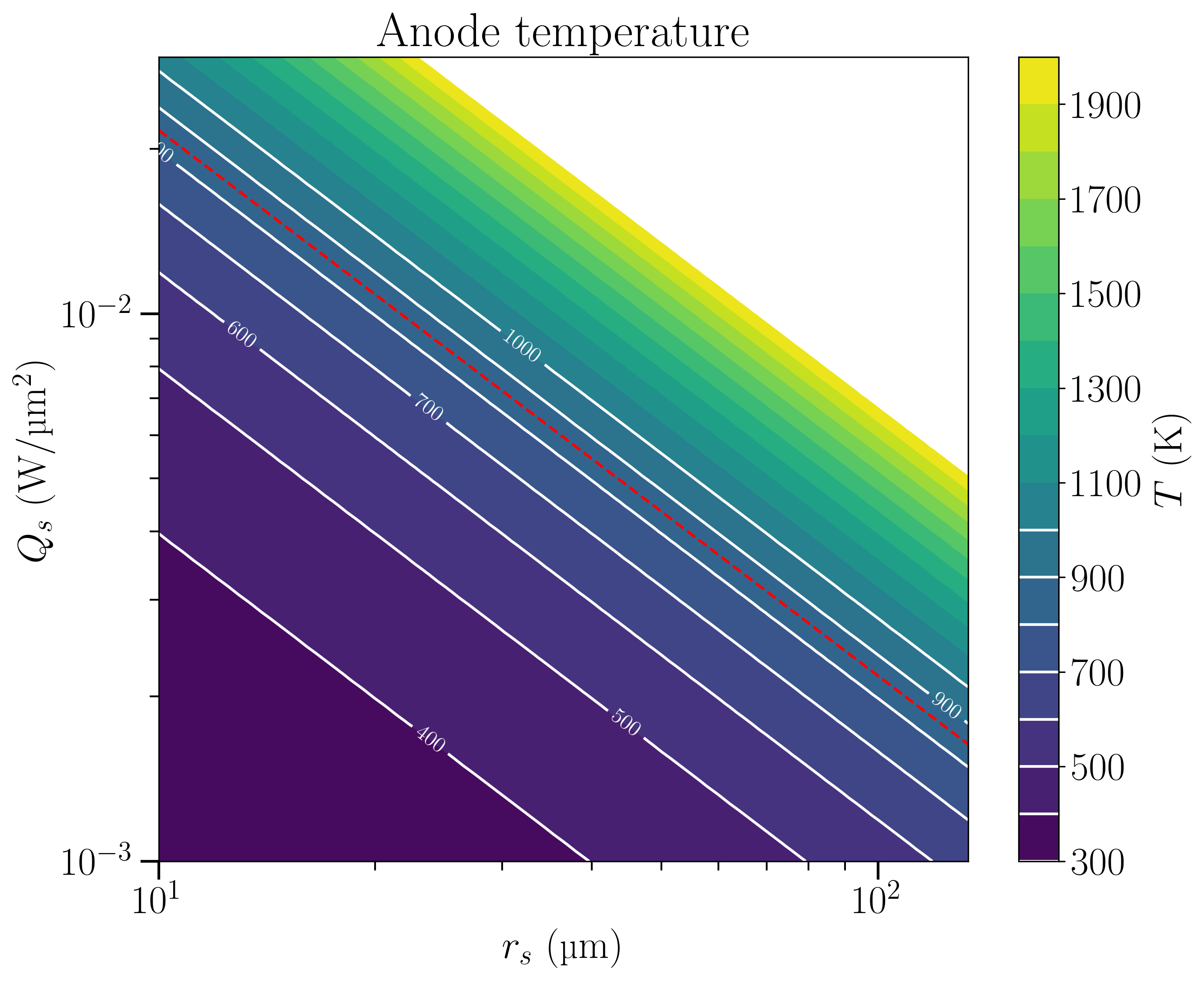}
 {Dependence of the maximum steady state temperature (up to 2000 K) on the Cu anode surface as a function of heat spot radius $r_s$ and surface power density $Q_s$ (energy transfer rate per unit area).\label{fig:tanode}}{0.95}

\section{Discussion}\label{sec:discussion}

The presence of high magnetic fields has been shown to decrease the performance of devices operating at high electric fields~\cite{mor05}. There is support for the theory that magnetic fields can significantly focus emitted electron beams and lead to the heating of metal surfaces at impact. Our simulations have demonstrated that the increase in temperature can be very significant, depending on voltage, magnetic field, or geometry. Localized temperatures can be high enough for metal evaporation to start, which is an important precursor process to vacuum arc formation. Sufficient evaporation can lead to ionization and the formation of a conducting pathway between the anode and cathode. These results suggest that arc formation may begin on the anode side when high magnetic fields are present.

Particle simulations suggest that the geometry of the metal surfaces, as well as the shape of the electric and magnetic fields, can significantly impact the motion of the emitted electron beam. Simulation of these processes can give insight beyond theoretical solutions, since actual accelerating structure geometries can be studied in detail. Depending on the situation, emitted electron beams can be either focused or defocused, which changes the heating dynamics of the anode surface. A more focused beam leads to higher current density and greater heat flux on the anode surface, leading to higher temperatures and possible evaporation of neutral atoms from the anode surface. The strength of the magnetic field is a major factor influencing heating of the anode surface and arc occurrence, since greater magnetic fields have the potential to lead to more tightly focused electron beams.

Previous simulations have considered fatigue damage and plastic deformation as the origin for arc formation~\cite{pal09, bow20}. We found that at local electric fields of 10~GV/m, emitted electron current can lead to very high temperatures on the anode side at magnetic fields on the order of 10--30~T. At such temperatures, metal evaporation flux on the anode surface would be high, suggesting an anodic arc formation mechanism. Surface temperatures will depend on the energy deposited by the electron beam, which results from the accelerating electric field, i.e. applied voltage and gap size. In pulsed mode, the heating is greatly reduced compared to the DC case.

We have focused on electron dynamics and heating, ignoring the formation of plasma that has been studied in cathodic arcing~\cite{tim15, koi24}. The role of the magnetic field for plasma formation is still unclear, but magnetic fields could lead to plasma confinement. This would lead to higher number densities, higher collision rates and possibly faster plasma development. Bombardment of the cathode by ions originating from the anode could lead to further plasma development on the cathode side. While the dynamics of nanotips has been studied previously~\cite{ves20}, the presence of an external magnetic field could impact this process. An external magnetic field could lead to additional forces and torque on an emitting tip. Additionally, the role of the material used can have an impact. The performance of materials such as Be, Cu and Al has been compared previously~\cite{bow20}, which suggests significant differences. To study this material dependence in more detail, a model considering both their thermal and mechanical properties would be needed.

\section{Conclusions}\label{sec:conclusions}

High magnetic fields can change the nature of vacuum arc occurrence and reduce the performance of devices operating at high electric fields. Therefore, understanding the vacuum arcing phenomenon in the presence of magnetic fields is important for the design of such devices. We have studied the behavior of emitted electron beams in high electric and magnetic fields to better understand the processes that contribute to arc occurrence. The direction of the electric and magnetic fields was shown to be an important factor for electron beam focusing and heating of the anode surface, which can contribute to plasma formation. The case where the electric and magnetic fields are in the parallel configuration was found to be most detrimental for performance, leading to the greatest anode heating, while the opposite was observed in the perpendicular case. Similarly, the magnitude of the magnetic field was found to be major contributor to the heating on the anode surface. The temperatures reached on the anode surface suggest that evaporation of neutrals from the anode side is a possibility in high magnetic fields. Our results support the role of focused electron beams as the origin of the anodic vacuum arcing mechanism. A direction for future work is the simulation of full anode-initiated plasma formation in the cathode-anode gap.


\section*{Acknowledgements}

RK, AK, TT and VZ are supported by the European Union's Horizon 2020 research and innovation program, under Grant Agreement No. 856705 (ERA Chair ``MATTER'') and by the Estonian Research Council's grants PRG2675, TARISTU24-TK10 and TEM-TA23. TT is supported by the Estonian Research Council Grant No. SJD61. RK, AK and FD have been supported by CERN CLIC K-contract (No. 47207461). RK and MT are supported by the doctoral program MATRENA of the University of Helsinki. The authors wish to thank the Finnish Computing Competence Infrastructure (FCCI) for supporting this project with computational and data storage resources.

\section*{Conflict of Interest}
The authors have no conflicts to disclose.

\section*{Author contributions}

\textbf{Roni Koitermaa:} Conceptualization, Data Curation, Formal Analysis, Investigation, Methodology, Software, Visualization, Writing - Original Draft Preparation, Writing - Review and Editing.
\textbf{Marzhan Toktaganova:} Conceptualization, Data Curation, Formal Analysis, Software, Visualization, Writing - Original Draft Preparation, Writing - Review and Editing.
\textbf{Andreas Kyritsakis:} 
Conceptualization, Funding Acquisition, Investigation, Project Administration, Resources, Software, Supervision, Writing - Review and Editing.
\textbf{Tauno Tiirats:} 
Conceptualization, Funding Acquisition, Investigation, Project Administration, Resources, Software, Supervision, Writing - Review and Editing.
\textbf{Alexej Grudiev:} 
Conceptualization, Writing - Review and Editing.   
\textbf{Veronika Zadin:} 
Conceptualization, Funding Acquisition, Investigation, Project Administration, Resources, Supervision, Writing - Review and Editing.
\textbf{Flyura Djurabekova:} 
Conceptualization, Funding Acquisition, Investigation, Project Administration, Resources, Supervision, Writing - Review and Editing.   

\section*{Data availability}

Data will be made available on request.

\bibliography{vacuum_arc_magnetic_field}

\end{document}